\RequirePackage{fix-cm}
\documentclass{svjour3}                     
\smartqed  
\usepackage{graphicx}
\usepackage{url}
\usepackage{natbib}
\usepackage{lineno}
\journalname{Journal}
\begin{document}

\title{Three-dimensional weights of evidence modeling of a deep-seated porphyry Cu deposit
}

\titlerunning{Three-dimensional weights of evidence modeling}        

\author{Ehsan Farahbakhsh \and
        Ardeshir Hezarkhani \and
        Taymour Eslamkish \and
        Abbas Bahroudi \and
        Rohitash Chandra
}


\institute{E. Farahbakhsh \at
              Department of Mining Engineering, Amirkabir University of Technology (Tehran Polytechnic), Tehran, Iran \\
              \email{e.farahbakhsh@aut.ac.ir} 
           \and
           A. Hezarkhani \at
              Department of Mining Engineering, Amirkabir University of Technology (Tehran Polytechnic), Tehran, Iran \\
           \and
           T. Eslamkish \at
              Department of Mining Engineering, Amirkabir University of Technology (Tehran Polytechnic), Tehran, Iran \\
           \and
           A. Bahroudi \at
              School of Mining Engineering, College of Engineering, University of Tehran, Tehran, Iran \\
           \and
           R. Chandra \at
              School of Mathematics and Statistics, University of New South Wales, Sydney, Australia \\
}

\date{Received: date / Accepted: date}

\maketitle

\begin{abstract}
Given the challenges in data acquisition and modeling at the stage of detailed exploration, it is difficult to develop a prospectivity model, particularly for disseminated ore deposits. Recently, the weights of evidence (WofE) method has demonstrated a high efficiency for modeling such deposits. In this study, we propose a framework for creating a three-dimensional (3D) weights of evidence-based prospectivity model of the Nochoun porphyry Cu deposit in the Urmia-Dokhtar magmatic arc of Iran. The input data include qualitative geological and quantitative geochemical information obtained from boreholes and field observations. We combine ordinary and fuzzy weights of evidence for integrating qualitative and quantitative exploration criteria in a 3D space constrained by a metallogenic model of the study area for identifying a deep-seated ore body. Ordinary weights of evidence are determined for geological data, including lithology, alteration, rock type, and structure. Moreover, we determine the fuzzy weight of evidence for each class of continuous geochemical models created based on Fe, Mo, and Zn concentration values derived from boreholes. We integrate the input evidential models using WofE and create two prospectivity models (i.e., posterior and studentized posterior probability). We also determine anomalous voxels in the probability models using concentration-volume fractal models and validate them using prediction-volume plots. The modeling results indicate that the studentized posterior probability model is more efficient in identifying voxels representing copper mineralized rock volumes. We provide open source software for the proposed framework which can be used for exploring deep-seated ore bodies in other regions.
\end{abstract}

\section*{Highlights}

\begin{itemize}
  \item Three-dimensional weights of evidence method is an efficient tool for modeling deep-seated ore bodies;
  \item Probability models are able to show the dispersion of target mineralization in depth;
  \item Studentized posterior probability model provides more reliable results compared to posterior probability model.
\end{itemize}

\section{Introduction}
\label{sec1}
The increasing shortage of easily detectable, outcropping ore deposits has led more and more mineral explorers to prospect for concealed or deep-seated ore deposits, in particular for metals such as copper, which plays a crucial role in modern society \citep{Mudd2013,Mudd2018,Schodde2013}. With increasing depth, traditional exploration methods are progressively becoming less efficient and/or more costly. Mineral prospectivity mapping has been developed and applied for various types of ore deposits and at a variety of scales ranging from continental to regional \citep{Brown2000,Carranza2005,Carranza2010,Chen2017,Rodriguez-Galiano2015,Xiong2018,Zuo2011}. Whilst most prospectivity mapping methods can be categorized as either knowledge- or data-driven approaches \citep{Cheng1999,Manap2013,Porwal2003}, hybrid methods consider both data and expert knowledge \citep{Sun2019}. The latter are typically used for identifying areas of high potential for the discovery of ore deposits in two-dimensional environments \citep{Carranza2008,Carranza2015,Knox-Robinson2000,Porwal2010}. \\
Deep-seated ore deposits usually show weak exploration signals on the ground surface. Therefore, there is a need to develop two-dimensional (2D) mineral prospectivity mapping methods in a three-dimensional (3D) space to benefit them for in-depth exploration of mineral resources. 3D modeling, analysis and visualization facilitate the perception of key spatial factors in mineralization, ore genesis, and geologic evolution in addition to target appraisal \citep{Carranza2009,Li2018,Mao2019,Payne2015,Zuo2016}. The ability of 3D modeling in providing a reliable spatial model is completely dependent on the quality of input datasets, modeling techniques, expert knowledge, and the complexity of the local geological setting \citep{Fallara2006,Houlding1994,Jessell2014,Lindsay2012,Liu2016}. A comprehensive metallogenic model such as those suggested for porphyry Cu deposits \citep{Berger2008,Lowell1970,Meng1997,Sillitoe2010}, helps geometric modeling and spatial analysis through enhancing the reliability of 3D models. \\
Several three-dimensional mineral prospectivity mapping methods have been developed in recent years \citep{Li2015,Mao2019,Nielsen2019,Nielsen2015,Xiao2015,Yuan2014}, which can be applied along with other modeling methods, such as geostatistics for modeling drilling data and detecting deep-seated ore deposits at both regional and local scales. Regional-scale 3D mineral prospectivity modeling and quantitative assessment is rarely feasible, because required public-domain datasets with consistent coverage over large areas are not available \citep{Xiao2015}. One of the advantages of 3D mineral prospectivity modeling over traditional geostatistics is its ability of integrating different types of qualitative and quantitative exploration data rather than being restricted to modeling the concentrations of individual geochemical elements. Some information, such as geological characteristics obtained from boreholes consume large amounts of time and money, which are surprisingly less considered at the stage of detailed exploration due to the lack of a specific framework for combining qualitative data with more favorable quantitative data. The 3D mineral prospectivity modeling methods are able to integrate such data and provide an efficient model for optimizing the process of selecting new drilling locations and planning the exploitation of an ore reserve. \\
In recent years, Bayesian modeling approaches have been applied to modeling geological features \citep{Olierook2019,Scalzo2019}. The Bayesian inference approach is able to provide a fully quantitative and informative 3D prospectivity model and to fuse all available constraints in a probabilistically rigorous fashion. The weights of evidence (WofE) method is based on Bayes' rule \citep{Xiao2015} and has been effectively used for 2D mineral prospectivity mapping of various types of ore deposits (e.g., \cite{Carranza2004,Kreuzer2015,Pazand2014,Porwal2010,Zeghouane2016}). The fuzzy WofE method developed by \cite{Cheng1999}, prevents loss of information due to converting continuous models into binary models. This method has a number of advantages compared to other simpler or even more complicated data-driven methods. In general, the weights of evidence represent the degree of correlation between a target mineralization and a particular model or pattern created under specific conditions known as an evidence \citep{Agterberg1990,Bonham-Carter1989,Carranza2004,Cheng1999,Yuan2014}. \\
Using 3D mineral prospectivity modeling, geological, geochemical and geophysical data can be integrated according to the known dispersion of mineralization in a modeling space. The result of this process is a formulated 3D model which presents a quantitative assessment of the probability of detecting a target mineralization based on exploration criteria. The exploration criteria must include all the factors which control a specific target mineralization in a study area \citep{Yuan2014}. They help to provide the 3D evidential models which are later used as inputs to the modeling process. \\
In this study, we extend and demonstrate the application of WofE method in a 3D space based on a proposed framework for modeling a porphyry copper (Cu) deposit located in southeast Iran within a magmatic arc called Urmia-Dokhtar. We use borehole data consisting of qualitative geological and quantitative geochemical data along with surficial geological data for providing input 3D evidential models. Based on the proposed framework, the ordinary and fuzzy WofE methods are used for weighting voxels in binary geological and continuous geochemical evidential models, respectively. We use the total variance associated with each voxel to create a studentized posterior probability model in addition to a posterior probability model. The concentration-volume fractal models and prediction-volume plots are used to evaluate and validate our models. We provide Python scripts as open-source software for implementing the proposed framework in this study.

\section{Geological setting}
\label{sec2}
The Nochoun porphyry copper deposit is located within a magmatic arc subdivision called Urmia-Dokhtar where extensive Tertiary to Plio-Quaternary intrusive and extrusive units are exposed in a northwest--southeast trend (Fig. 1). In several studies, a subduction-related magmatic model is suggested for the Urmia-Dokhtar magmatic arc, which is known to be a result of the closure of the Neo-Tethys ocean between Arabian and Eurasian plates \citep{Berberian1981,Omrani2008}. In general, this magmatic arc involves two major mineralization regions of Chahar Gonbad to the southeast and Sungun to the northwest. The dominant type of mineralization is porphyry Cu which is associated with Eocene, Pliocene and Quaternary plutonic bodies and volcanic rocks. The major lithological units of the study area located in the Chahar Gonbad region in southeast Iran, consist of volcanic and subvolcanic complexes, and intrusive bodies. The volcanic rocks cover most of the study area and consist of Eocene andesite, dacite, and rhyodacite associated with tuff breccias \citep{Abedi2014}. As shown in Fig. 2, the intrusive bodies include granite to diorite dispersed in the south to southwest of the study area.

\begin{figure}
  \centering
  \frame{\includegraphics[width=\linewidth]{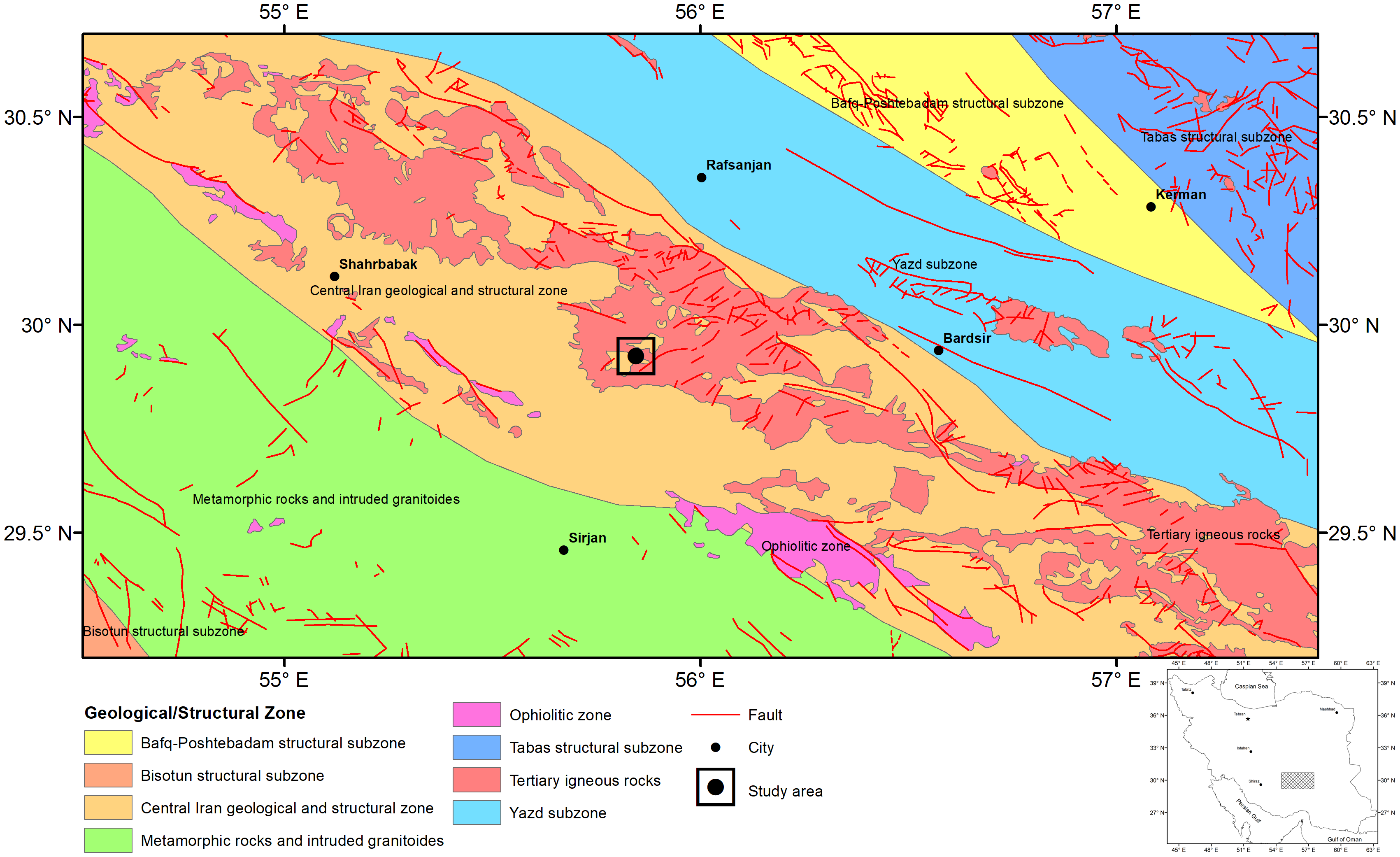}}
  \caption{A portion of the Chahar Gonbad region located in southeast Iran within the Urmia-Dokhtar magmatic arc. The study area is shown at the center of the map in a black square.}
  \label{fig1}
\end{figure}

\begin{figure}
  \centering
  \frame{\includegraphics[width=\linewidth]{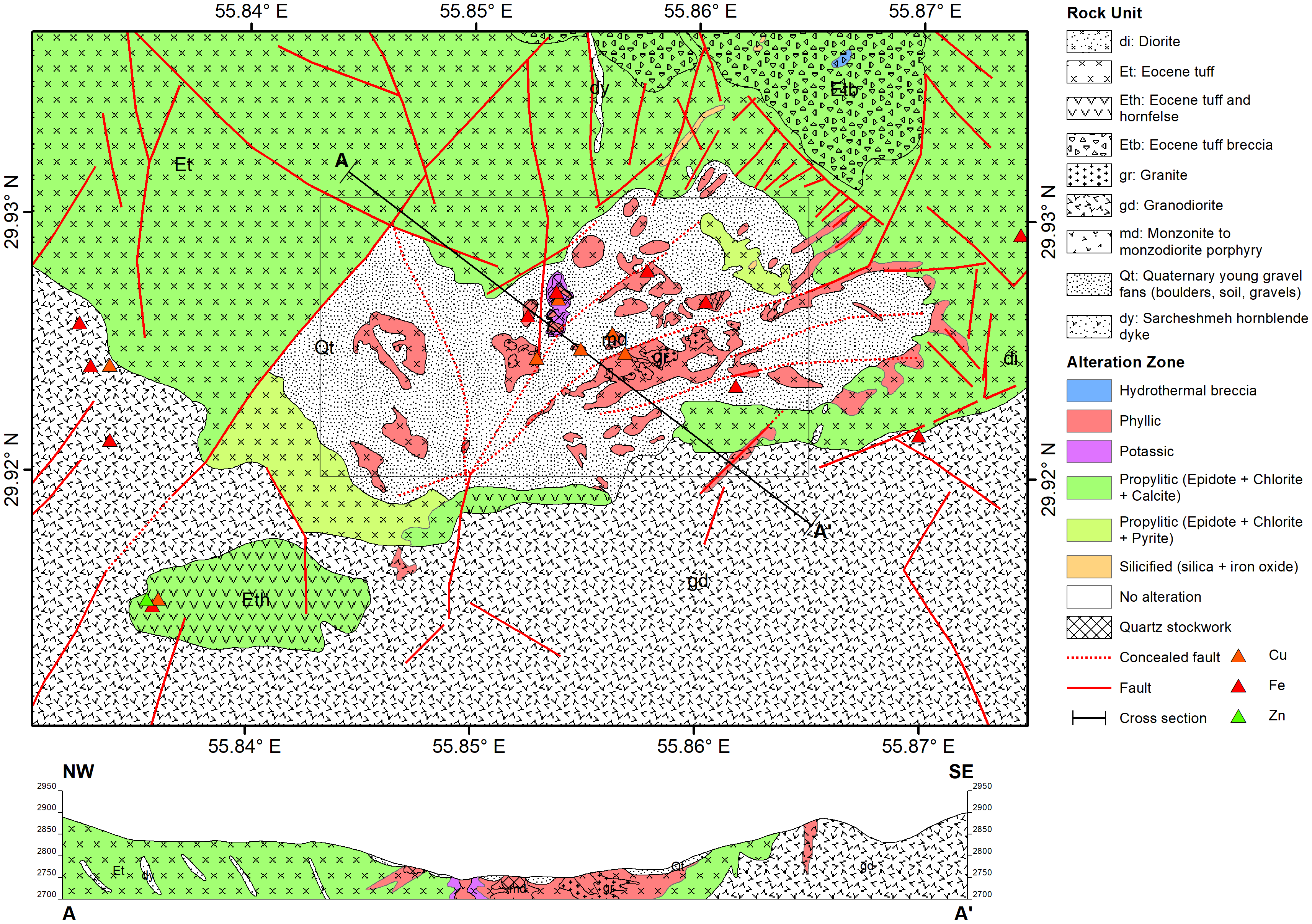}}
  \caption{Detailed geological and alteration map of the study area along with a cross-section intersecting some of the mineral occurrences. The black rectangle shows the target area.}
  \label{fig2}
\end{figure}

\section{Alteration and mineralization}
\label{sec3}
The hydrothermal alteration and mineralization in the Nochoun porphyry copper deposit strike NE--SW and center on a stock (Fig. 2). The early hydrothermal alteration was dominantly potassic and propylitic, followed later by phyllic, silica, and argillic alteration types. The potassic alteration is represented by mineral assemblages developed pervasively as halos around veins in the central parts of the study area. There is a relatively sharp boundary between the propylitic and potassic alteration zones in deeper parts of the ore deposit, but this contact is obscured by later phyllic alteration in shallow levels. The propylitic alteration is pervasive and represented mainly by chloritization of primary and secondary biotite and groundmass materials of the rocks which are peripheral to the central potassic zone (Fig. 2). The plagioclase minerals were replaced with epidote, but this alteration is less pervasive and less intense compared to the chloritization. Feldspar minerals are locally altered to clay minerals in shallow levels and the dominant mineral is kaolinite accompanied by illite. Moreover, the entire rocks have been altered to an assemblage of clay minerals, hematite, and quartz close to the erosion surface which are soft and white. The shallow alteration is interpreted to represent a supergene blanket over the ore deposit and the alteration of feldspar to clay in depth may have the same origin. Also, the latter may represent an argillic stage of the hypogene alteration. \\
The hypogene copper mineralization has disseminated or taken place in veinlets during phyllic alteration and to a lesser extent potassic alteration. During the potassic alteration, the copper was deposited as chalcopyrite and minor bornite; later hypogene copper was deposited mainly as chalcopyrite. The rocks are highly altered at the exposed surface of the ore deposit and the only mineral which has survived alteration is quartz. Most of the sulfide minerals have been leached, and copper has been concentrated in an underlying supergene zone by downward percolating groundwater. In general, the mineralization in the study area is the result of two geological processes including the intrusion of a granodiorite body and the ascending of hydrothermal fluids. The mineralization-bearing rocks hosted by the granodiorite body usually appear in the veins with a thickness of 2--3 m, and they are mostly found in the marginal sections of the host body or tuffaceous units. This type of mineralization is considered non-economic due to the low grade. Different types of ore minerals such as hematite, oligist, magnetite, and malachite with intense silicification of the veins are observed in mineralization zones (Fig. 3). Moreover, the polished sections provided using mineralized rocks, show the presence of chalcopyrite, pyrite, and sphalerite (Fig. 4). There are a number of stockworks exposed on the surface which are the result of hydrothermal fluid interaction (Fig. 2). These stockworks are probably related to the intrusion of dykes and quartz monzonite apophyses. In this type of mineralization, malachite and azurite minerals are found on the surface along with minor chalcopyrite as inclusions within quartz \citep{Abedi2014}.

\begin{figure}
  \centering
  \includegraphics[width=\linewidth]{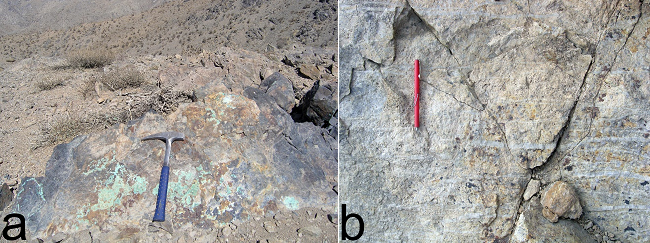}
  \caption{a) Cu oxide minerals including malachite and azurite in fractures of silicified tuff crystals; b) quartz veinlets with a stockwork texture.}
  \label{fig3}
\end{figure}

\begin{figure}
  \centering
  \includegraphics[width=\linewidth]{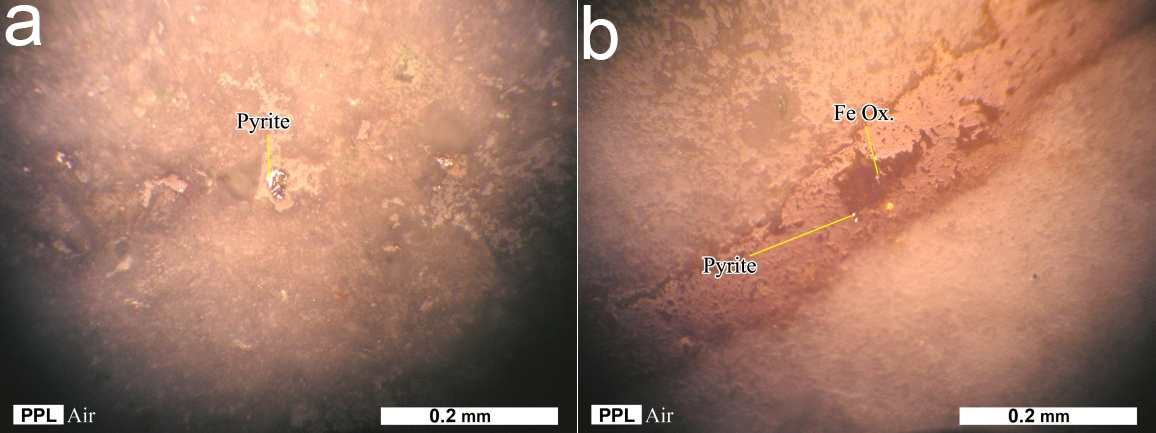}
  \caption{Polished sections showing the presence of a) pyrite, and b) iron oxide crystals in mineralized rocks.}
  \label{fig4}
\end{figure}

\section{Materials and methods}
\label{sec4}
\subsection{Drilling data}
\label{sec4-1}
We use qualitative geological and quantitative geochemical data obtained from 113 boreholes to create 3D geological and geochemical evidential models. The size of each voxel in the 3D models according to the extent of the modeling space is determined 10$\times$10$\times$10 meters (m). The geometrical parameters of the modeling space can be found in Table 1. We inscribe the 3D models in a polygon created by the convex hull algorithm based on the coordinates of the borehole collars on the ground surface (Fig. 5). Moreover, they are restricted to a super- and sub-face based on the elevation of the borehole collars and the depth of each borehole to make sure there are sufficient number of data points for interpolating throughout the modeling space. The total number of voxels is approximately 500,000. \\

\begin{table}[]
  \caption{Geometrical parameters of the 3D modeling space.}
  \label{table1}
  \begin{tabular}{ll}
    \hline\noalign{\smallskip}
    Parameter                   & Value      \\
    \noalign{\smallskip}\hline\noalign{\smallskip}
    North--South extent          & 970 m      \\
    East--West extent            & 1,740 m     \\
    Vertical extent             & 890 m      \\
    Polygon area on the surface & 0.8841 km\textsuperscript{2} \\
    \noalign{\smallskip}\hline
  \end{tabular}
\end{table}

The geological data used for creating 3D geological evidential models involve lithology, alteration, and rock type information. In addition to Cu concentration values used for creating a primary model of the target ore body, we use other geochemical data including the concentration values of key elements for creating 3D geochemical evidential models. The 3D strip-logs of the geological and geochemical data are shown in Figs. 6 and 7, respectively.

\begin{figure}
  \centering
  \frame{\includegraphics[width=\linewidth]{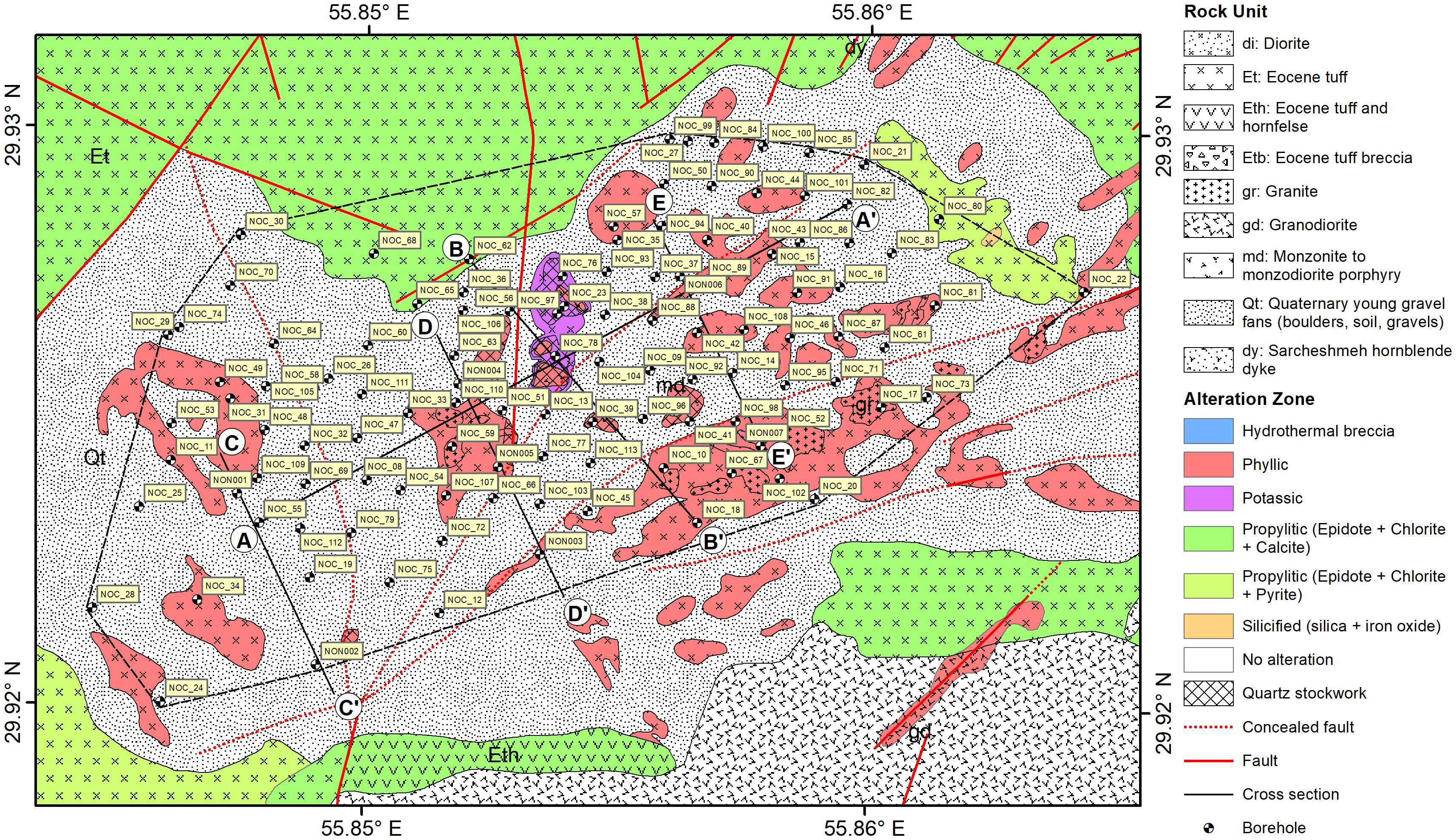}}
  \caption{Borehole collars along with their identification labels shown on the geological map of the study area enlarged from Fig. 2. The polygon drawn around the borehole collars shows the boundaries of the modeling space.}
  \label{fig5}
\end{figure}

\begin{figure}
  \centering
  \frame{\includegraphics[width=0.7\linewidth]{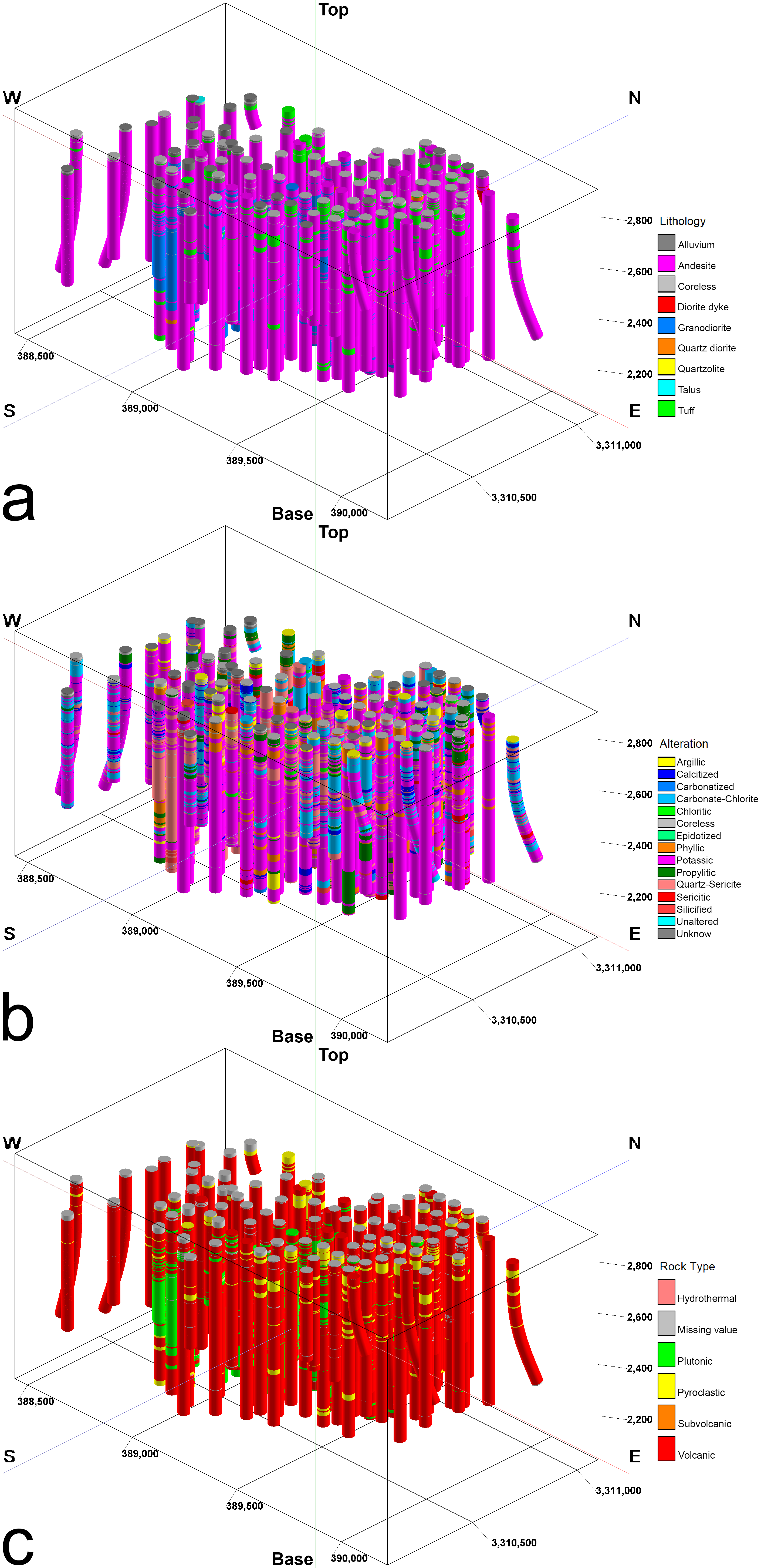}}
  \caption{3D strip-logs of the geological data including a) lithology, b) alteration, and c) rock type.}
  \label{fig6}
\end{figure}

\begin{figure}
  \centering
  \frame{\includegraphics[width=\linewidth]{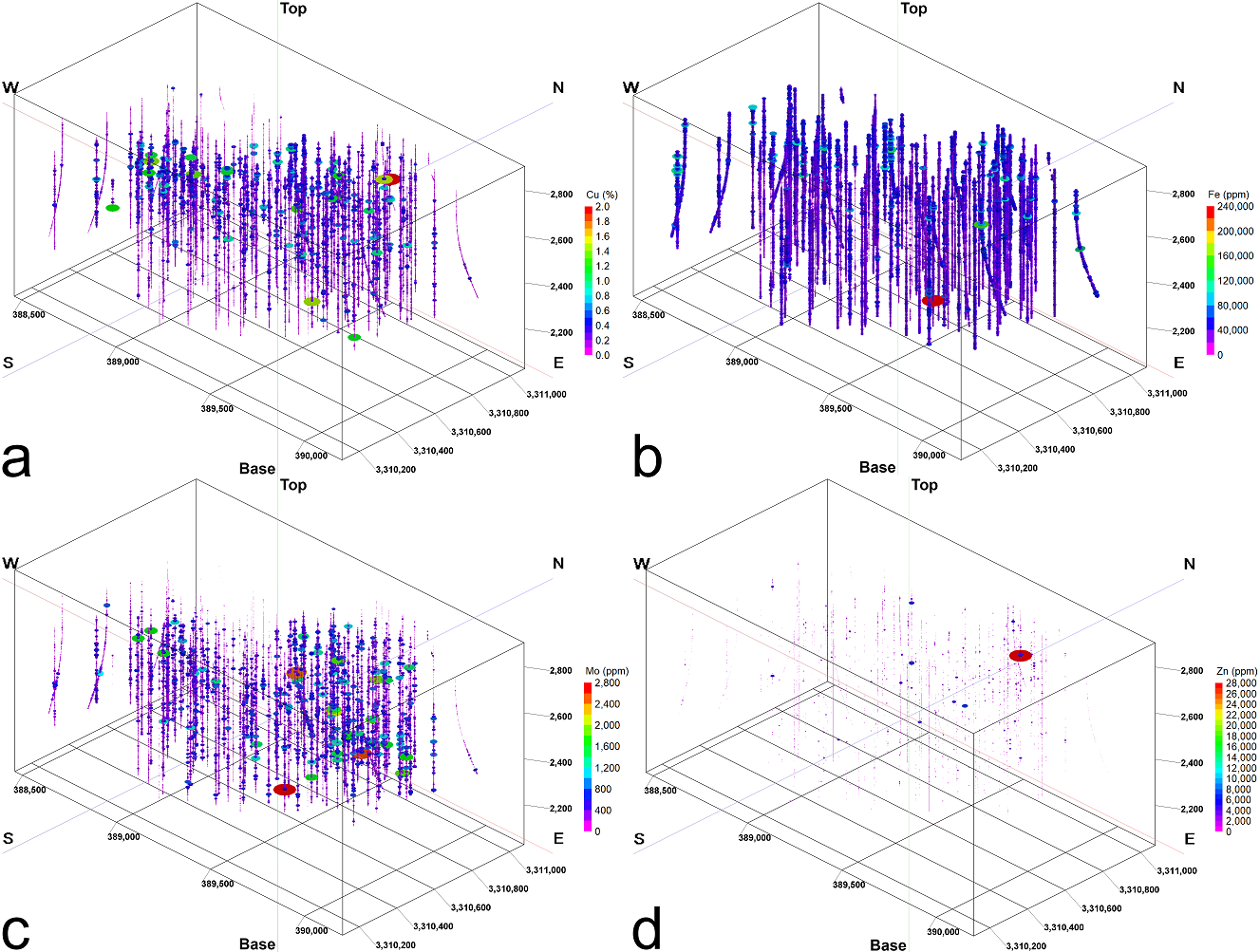}}
  \caption{3D strip-logs of the geochemical data including a) Cu, b) Fe, c) Mo, and d) Zn concentration values. The size and color of cylinders shown along the boreholes are proportionate to the intensity of concentration values. Highly positive skewed Zn concentration values yields a poor illustration of the relevant 3D strip-logs.}
  \label{fig7}
\end{figure}

\subsection{Structural data}
\label{sec4-2}
The drilling data used in this study do not involve structural data such as the position of faults in depth. Therefore, we extend surficial structural data to depth for creating 3D fault surfaces. The available surficial data include some information about the strike, dip, and dip direction of the faults which have been provided by the fieldworks, and validated by remote sensing data \citep{Farahbakhsh2019a}. The faults include those exposed on the surface or covered by a thin layer of regolith known as concealed faults. A 3D model of the fault surfaces called 3D ribbons is created using the available data which is shown in Fig. 8. The 3D ribbons are then restricted to the modeling space and converted into a 3D block model to be used as a structural evidential model.

\begin{figure}
  \centering
  \frame{\includegraphics[width=\linewidth]{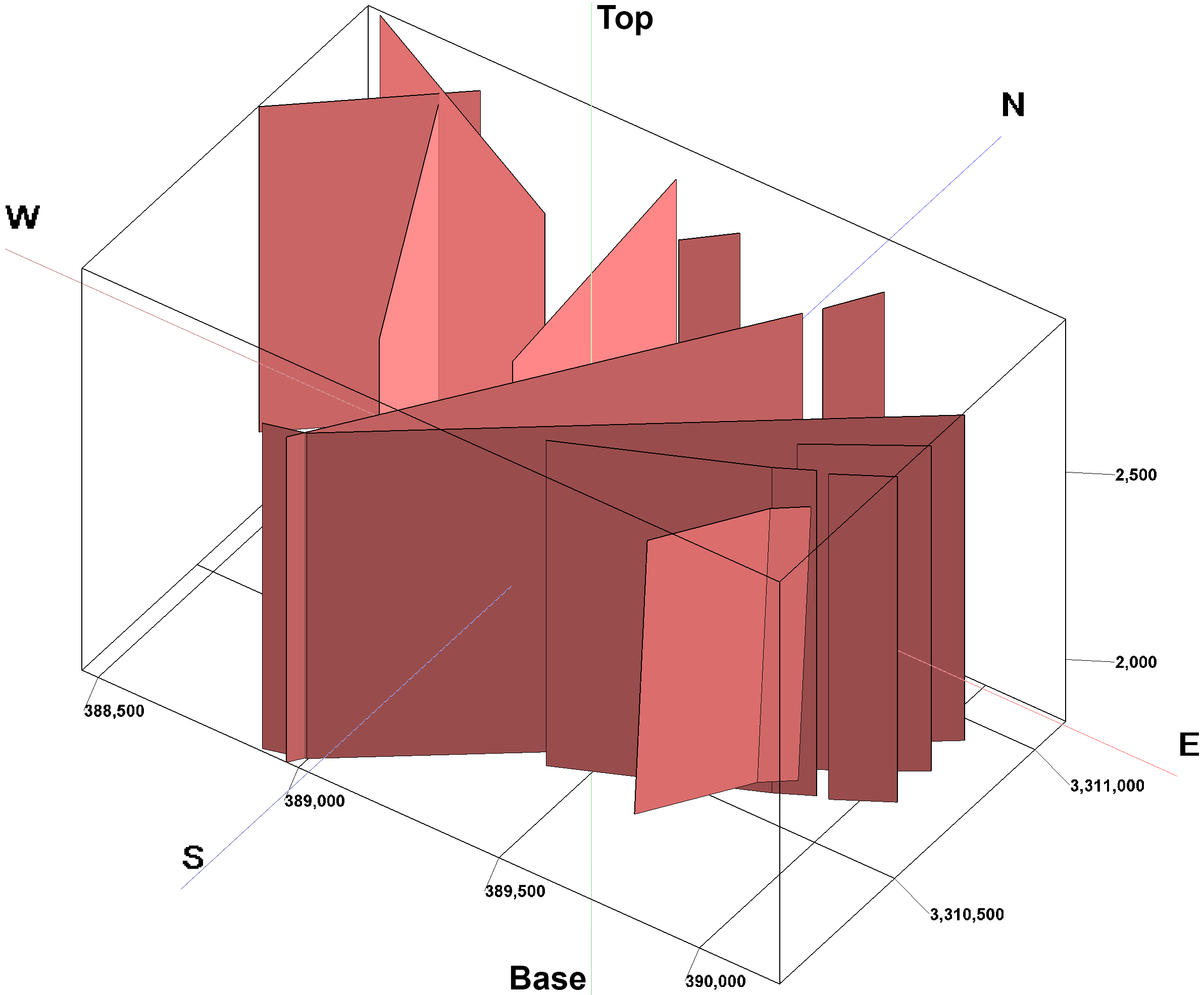}}
  \caption{3D ribbon model showing the fault surfaces which have been created using surficial structural data.}
  \label{fig8}
\end{figure}

\subsection{3D modeling}
\label{sec4-3}
\subsubsection{Geological modeling}
\label{sec4-3-1}
As shown in Fig. 5, there are a high number of boreholes in a small area which makes it possible to create a precise 3D geological model using the drilling data. In this study, we use a basic interpolation method called closest point for interpolating qualitative geological data including lithology, alteration, and rock type in a 3D space with the RockWorks software package \citep{RockWorks172019}. The value of a voxel node is set to be equal to the value of the nearest data point, regardless of its distance from the point or the value of its other neighbors. One of the advantages of this method is that the solid model nodes will honor the control points. This method can be used for modeling complex non-stratiform geology (e.g., multiple intrusions, impact craters, karst, etc.) such as our study area. The 3D geological evidential models are constrained to the geological map presented in Fig. 5. Moreover, the lithological model is constrained to the cross-sections shown in Fig. 9 drawn based on the drilling data and geological knowledge.

\begin{figure}
  \centering
  \frame{\includegraphics[width=\linewidth]{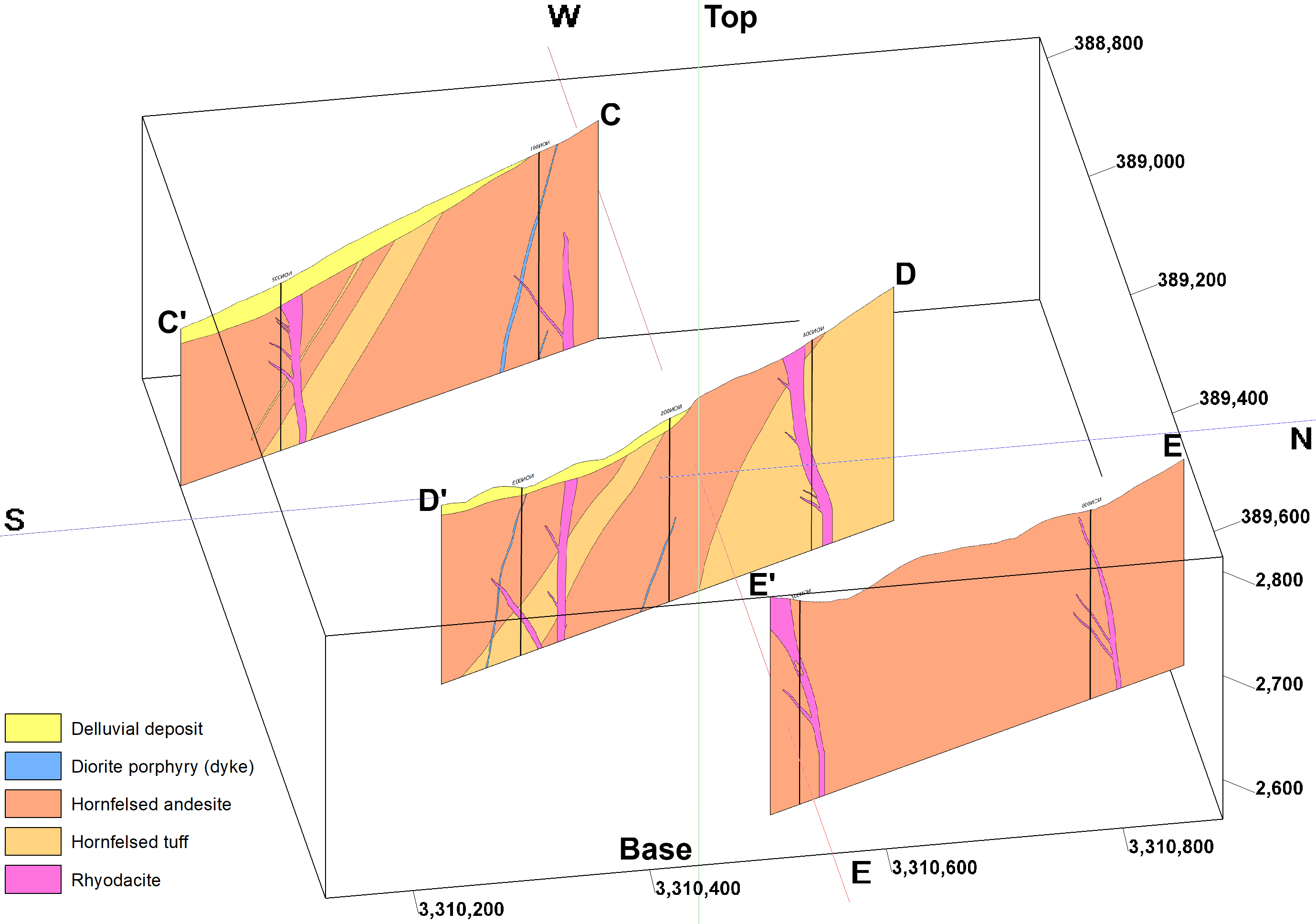}}
  \caption{Lithological cross-sections presented in a 3D view. The 3D lithological evidential model is constrained to these sections, in addition to the geological map.}
  \label{fig9}
\end{figure}

\subsubsection{Geochemical modeling}
\label{sec4-3-2}
The concentration values of different elements have been determined along the boreholes. However, the number of samples are different. We investigate the bivariate correlation between different elements and Cu concentration values. According to the results, iron, molybdenum, and zinc (Fe, Mo, and Zn) show a high correlation and they involve the highest number of samples. These elements are usually applied through prospecting for porphyry Cu deposits \citep{Farahbakhsh2019,Xiao2014}. Moreover, these elements are known as the key indications of porphyry Cu mineralization in the study area. We use the inverse-distance anisotropic modeling method which is one of the different kinds of the inverse-distance algorithm for interpolating the geochemical data in a 3D space. This method has been widely used for interpolating the concentration values of different geochemical elements in both two- and three-dimensional models, particularly in porphyry copper deposits (e.g., \cite{Wang2012,Zuo2011a}). \\
Using the inverse-distance in general, we assign a voxel node value based on the weighted average of neighboring data points, and the value of each data point is weighted according to the inverse of its distance from the voxel node, taken to a power. The greater the value of the exponent, the less influence distant control points will have on the assignment of the voxel node value \citep{RockWorks172019a,Zeghouane2016}. Using the inverse-distance anisotropic method, we look for the closest control point in each 90-degree sector around the node. In this study, the weighting exponent is set to 2, experimentally. The directional search can improve the interpolation of voxel values that lie between data point clusters, and can be useful for modeling borehole-based data. The quadrant searching tends to connect the limits (highs and lows) at the same elevation.

\subsection{Weights of evidence modeling}
\label{sec4-4}
The weights of evidence method is a well-known and robust data-driven prospectivity modeling method based on the Bayesian theorem \citep{Bonham-Carter1994}. This method is used for estimating the posterior probability of detecting an ore body under the assumption of conditional independence of input evidential models \citep{Xiao2015}. This assumption is also known as one of the weaknesses of this method \citep{Joly2012}. The WofE method can work with a low number of training datasets compared to other data-driven or machine learning techniques. Using this method, the prior belief of detecting a specific type of mineralization in an area or space is updated in the light of other evidence such as geological, geochemical or geophysical models. An example of the prior belief is the primary model created by interpolating the concentration values of target element obtained through boreholes. In this study, we interpolate the Cu concentration values as well as other geochemical elements, and then convert into a binary model for determining the approximate target ore body. In general, the posterior probability of mineralization (P(M)) after looking at evidence (E) is determined via the likelihood function using Eq. 1 \citep{Bonham-Carter1994}.

\begin{equation}
\label{eq1}
  P(M|E)=P(M)\frac{P(E|M)}{P(E)}
\end{equation}

This method is simple from the computational view, and 2D WofE method is readily implemented using GIS packages (e.g., \cite{ArcGISDesktop2019,QGISDevelopmentTeam2019}); however, implementing this method in a 3D space is more complicated. In this study, we deal with binary or discrete geological and continuous geochemical evidential models. Therefore, to minimize the loss of information, we determine ordinary positive and negative weights for each unit of the geological models and a fuzzy weight for each class of the geochemical models. In binary models, the weights of evidence enable a user to interpret positive and negative weights in geological terms intuitively. The positive (W\textsuperscript{+}) and negative (W\textsuperscript{-}) weights are determined using Eqs. 2 and 3 \citep{Bonham-Carter1994}.

\begin{equation}
\label{eq2}
  W^+=ln\frac{P(E|M)}{P(E|\bar{M})}
\end{equation}

\begin{equation}
\label{eq3}
  W^-=ln\frac{P(\bar{E}|M)}{P(\bar{E}|\bar{M})}
\end{equation}

These weights indicate the spatial association between the voxels with and without mineralization, and the presence and absence of anomalous voxels in evidential models, respectively. For example, an evidential or predictor model can be used to assess the contribution of a geological process in the formation and prospectivity of a specific type of mineralization \citep{Xiao2015}. Investigating the contrast (C) which is determined using Eq. 4, can help through this assessment \citep{Bonham-Carter1994}.

\begin{equation}
\label{eq4}
  C=W^+-W^-
\end{equation}

High values of contrast show strong association of an evidential model with the mineralization process. The contrast is used to calculate the fuzzy weight of each class in continuous models. In this study, we classify geochemical evidential models based on the percentile to minimize the effect of number of voxels on the fuzzy weight. Each model has ten classes with an equal number of voxels and a fuzzy membership function is created for each set of contrasts for different continuous evidential models. The contrast values are transformed to the fuzzy space ranging from 0 to 1 using a logistic function \citep{Yousefi2016} and called fuzzy contrast ($\mu_{E}(C)$). The fuzzy weight ($W_{\mu_{E}(C)}$) for each class of continuous models is determined using Eq. 5 \citep{Cheng1999}.

\begin{equation}
\label{eq5}
  W_{\mu_E(C)}=ln\frac{\mu_E(C)P(E|M)+(1-\mu_E(C))P(\bar{E}|M)}{\mu_E(C)P(E|\bar{M})+(1-\mu_E(C))P(\bar{E}|\bar{M})}
\end{equation}

The posterior probability model is the result of integrating input evidential models, which is generated using Bayes' equation in a log-linear form (Eq. 6) with the assumption that conditional independence applies \citep{Bonham-Carter1994,Bonham-Carter1989}.

\begin{equation}
\label{eq6}
  P_{Pst}=\frac{O_{Pst}}{1+O_{Pst}}
\end{equation}

Where, P\textsubscript{Pst} denotes the posterior probability and O\textsubscript{Pst} is the posterior odds. The posterior odds equal the exponent of the posterior logit which can be determined by Eq. 7 \citep{Bonham-Carter1994,Cheng1999}.

\begin{equation}
\label{eq7}
  lnO(M|E_1E_2 \ldots E_k)=lnO(M)+\sum_{i=1}^{m}W_i^{+/-}+\sum_{j=1}^{n}W_{\mu_j}
\end{equation}

Where, there are k evidential models including m discrete and n continuous models. We use the aggregate of positive or negative weights in discrete models and the fuzzy weights in continuous models, with the prior logit of mineralization in the study area for calculating the posterior logit. \\
Using the WofE method enables the user to calculate the effects of uncertainty on the weights, and uncertainty due to missing information. This leads to producing an uncertainty quantified model which is propagated in the decision making. The variances of the weights and contrast help to model the uncertainty of the posterior probability due to uncertainty in the weights and caused by lack of information. The variance of positive ($\sigma^2_{W^+}$) and negative ($\sigma^2_{W^-}$) weights in discrete models are determined as presented in Eqs. 8 and 9, respectively \citep{Bonham-Carter1994}.

\begin{equation}
\label{eq8}
  \sigma_{W^+}^2=\frac{1}{N(E \cap M)}+\frac{1}{N(E \cap \bar{M})}
\end{equation}

\begin{equation}
\label{eq9}
  \sigma_{W^-}^2=\frac{1}{N(\bar{E} \cap M)}+\frac{1}{N(\bar{E} \cap \bar{M})}
\end{equation}

Where, N is the number of voxels and the variance resulting from the membership function $\mu_E$ can be expressed as Eq. 10. It has to be noted that we have assumed $P(E)+P(\bar{E})=1$.

\begin{equation}
\label{eq10}
  \sigma_{\mu_E}^2[P(M)]=\frac{2\mu_E(1-\mu_E)}{P(\mu_E)}\sigma^2[P(M)]
\end{equation}

Where, $P(\mu_E)$ denotes the probability of the membership function and determined using Eq. 11. Also, $\sigma^2[P(M)]$ is the variance of the prior probability of mineralization as given by Eq. 12.

\begin{equation}
\label{eq11}
  P(\mu_E)=\mu_EP(E)+(1-\mu_E)P(\bar{E})
\end{equation}

\begin{equation}
\label{eq12}
  \sigma^2[P(M)]=\{P(M|E)-P(M)\}^2P(E)+\{P(M|\bar{E})-P(M)\}^2P(\bar{E})
\end{equation}

A useful measure is to calculate the studentized value of contrast (C\textsubscript{St}), as a measure of the uncertainty with which the contrast is known. We calculate the studentized value as the ratio of contrast to its standard deviation S(C) as shown in Eq. 13.

\begin{equation}
\label{eq13}
  C_{St}=\frac{C}{S(C)}
\end{equation}

The standard deviation of contrast is determined using Eq. 14.

\begin{equation}
\label{eq14}
  S(C)=\sqrt{\sigma_{W^+}^2+\sigma_{W^-}^2}
\end{equation}

A large studentized contrast implies that the contrast is large compared with the standard deviation, then the contrast is more likely to be real. A studentized value larger than 2, or even 1.5 is satisfactory \citep{Bonham-Carter1994}. Due to the assumptions required for a formal statistical test, particularly the problem with the dependence of the standard deviation of contrast on the units of measurement, it is best to use this ratio in a relative, rather than an absolute sense \citep{Bonham-Carter1994}. \\
We use the variances of the weights to calculate the variance of the posterior probability at each voxel, and to generate a studentized posterior probability model. The square root of total variance ($\sigma^2_T$) at each voxel equals the standard deviation. The ratio of the posterior probability to the corresponding standard deviation is called the studentized posterior probability (P\textsubscript{St}) as shown in Eq. 15.

\begin{equation}
\label{eq14}
  P_{St}=\frac{P_{Pst}}{\sqrt{\sigma_T^2}}=\frac{P_{Pst}}{\sqrt{\sum\sigma_{W^+}^2+\sum\sigma_{W^-}^2+\sum\sigma_{\mu_E}^2}}
\end{equation}

According to this equation, the ratio of the posterior probability to the square root of variances of positive, negative and fuzzy weights in both discrete and continuous models equals the studentized posterior probability in each voxel. The studentized posterior probability acts as a measure of the relative certainty of the posterior probability. The voxels where the studentized value falls below some threshold can be masked out, due to lack of confidence in the results. In this study, the threshold is determined based on the concentration-volume (C-V) fractal model \citep{Afzal2011}. \\
As described above, the three-dimensional WofE method used in this study is similar to the traditional method applied for mineral prospectivity mapping \citep{Bonham-Carter1994,Carranza2004,Cheng1999}, and pixels are replaced with voxels. We propose a framework for implementing the method in a 3D space and summarize the steps to be taken for 3D mineral potential modeling using the WofE method in Fig. 10. A brief description of these steps is presented in the following.

\begin{itemize}
  \item Acquiring data including required geological and geochemical data;
  \item Creating a primary model of the target mineralization based on the Cu concentration values obtained along the boreholes in different intervals;
  \item Providing 3D geological and geochemical evidential models;
  \item Calculating the weights of evidence and other necessary parameters such as the standard deviation and variance of the weights for different geological units and each class of geochemical models;
  \item Selecting input evidential models for creating the posterior probability model based on the contrast and studentized contrast;
  \item Integrating selected evidential models and creating 3D posterior probability and studentized posterior probability models;
  \item Validating the results.
\end{itemize}

\begin{figure}
  \centering
  \includegraphics[width=0.7\linewidth]{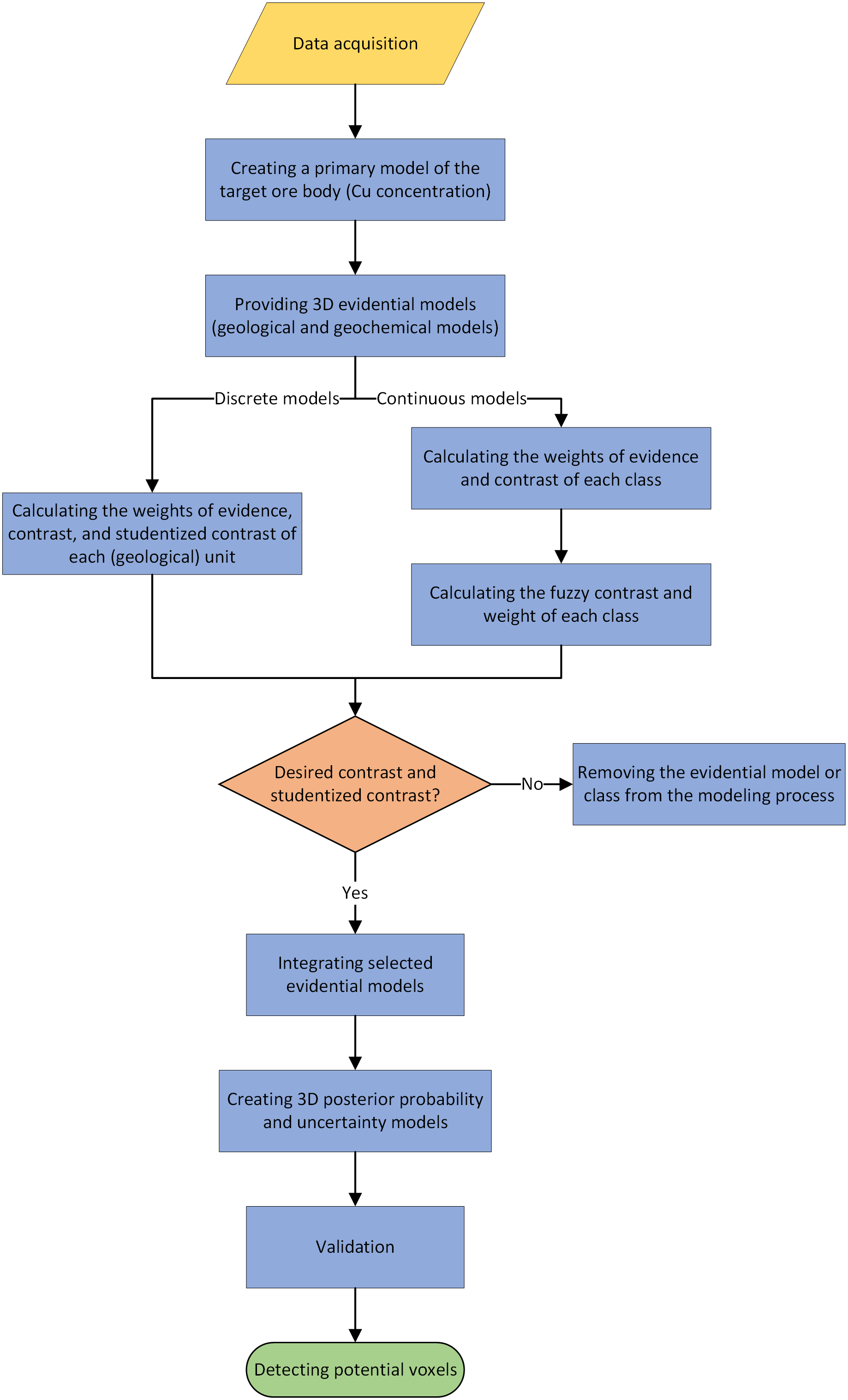}
  \caption{Methodology flowchart of this study for 3D mineral prospectivity modeling of an ore deposit using the WofE method.}
  \label{fig10}
\end{figure}

\section{Results}
\label{sec5}
\subsection{3D evidential models}
\label{sec5-1}
We use three different types of geological data including lithology, alteration, and rock type for creating 3D geological evidential models constrained to the geological map. As shown in Fig. 5, we considered two cross-sections ($AA'$ and $BB'$) along the small and large diameters of the study area to visually investigate the correlation between the aforementioned data types and the Cu concentration. These cross-sections along with the interpolated Cu concentration anomaly zones (greater than 0.2\%) are shown in Fig. 11. \\

\begin{figure}
  \centering
  \frame{\includegraphics[width=\linewidth]{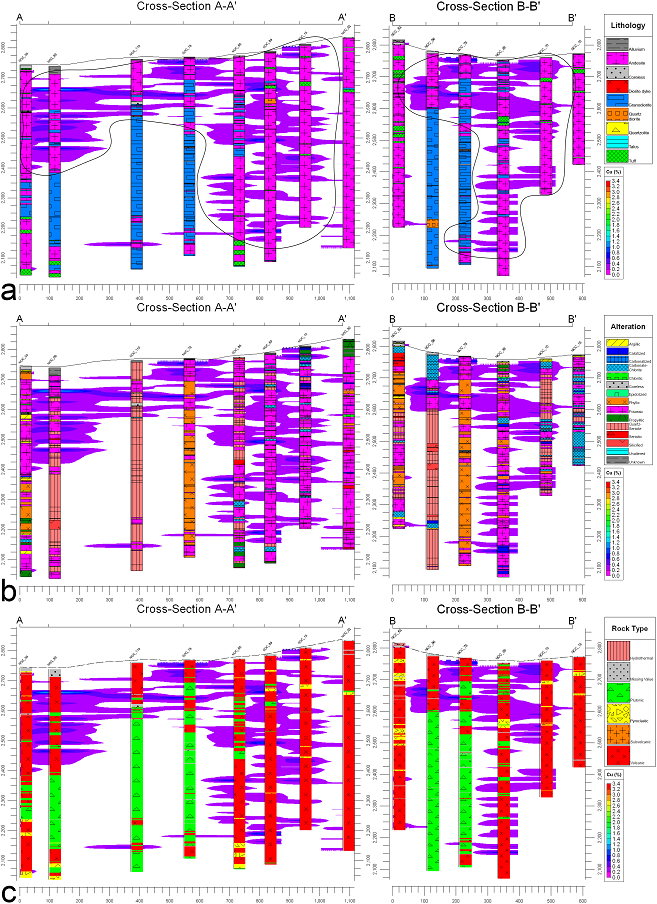}}
  \caption{Cross-sections $AA'$ and $BB'$ (Fig. 5) illustrated using a) lithology, b) alteration, and c) rock type data obtained from the boreholes. The Cu concentration anomaly zones are shown in the background and outlined using the black lines.}
  \label{fig11}
\end{figure}

As mentioned earlier, we used RockWorks software package \citep{RockWorks172019} for creating 3D solid models using different available geological data obtained from the boreholes. As described in section 4.3.1, according to the high density of the data points, a method called the closest point is used to interpolate the geological data in the 3D space. The geological models are constrained to the geological map (Fig. 5). Moreover, the lithology model is constrained to the cross-sections shown in Fig. 8 for providing a more precise and reliable model. The solid models created using lithology, alteration, and rock type data are presented in Fig. 12. \\

\begin{figure}
  \centering
  \frame{\includegraphics[width=0.7\linewidth]{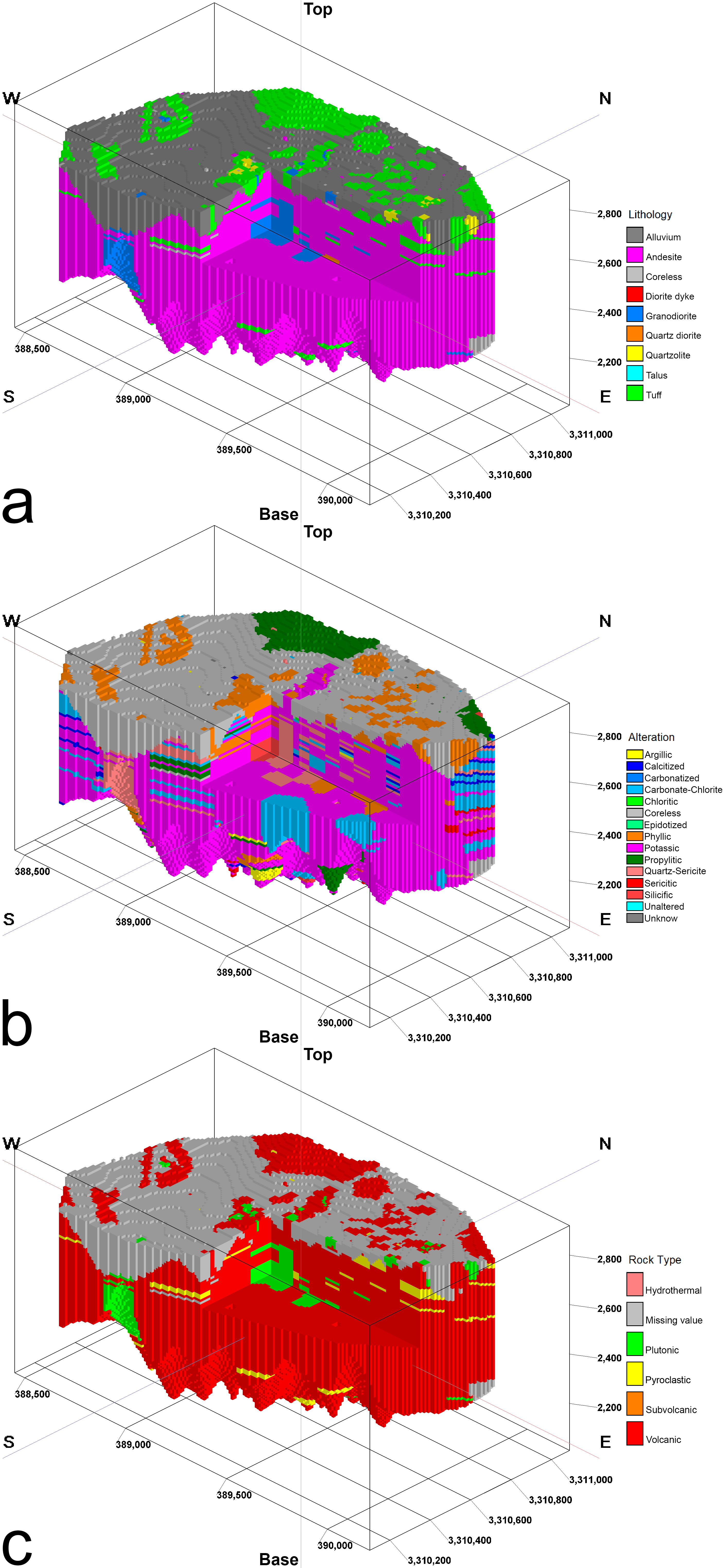}}
  \caption{Solid models created using a) lithology, b) alteration, and c) rock type data. Each solid model is shown with a cutout.}
  \label{fig12}
\end{figure}

The 3D ribbon model of the faults shown in Fig. 8 is converted into a block model. The size of each voxel in this model is the same as other evidential models. Moreover, we use two buffer zones with a radius of 25 and 50 m surrounding the fault blocks to investigate the correlation of proximity to the faults and Cu mineralization. An overall model of the fault blocks and different buffer zones is presented in Fig. 13. \\

\begin{figure}
  \centering
  \frame{\includegraphics[width=\linewidth]{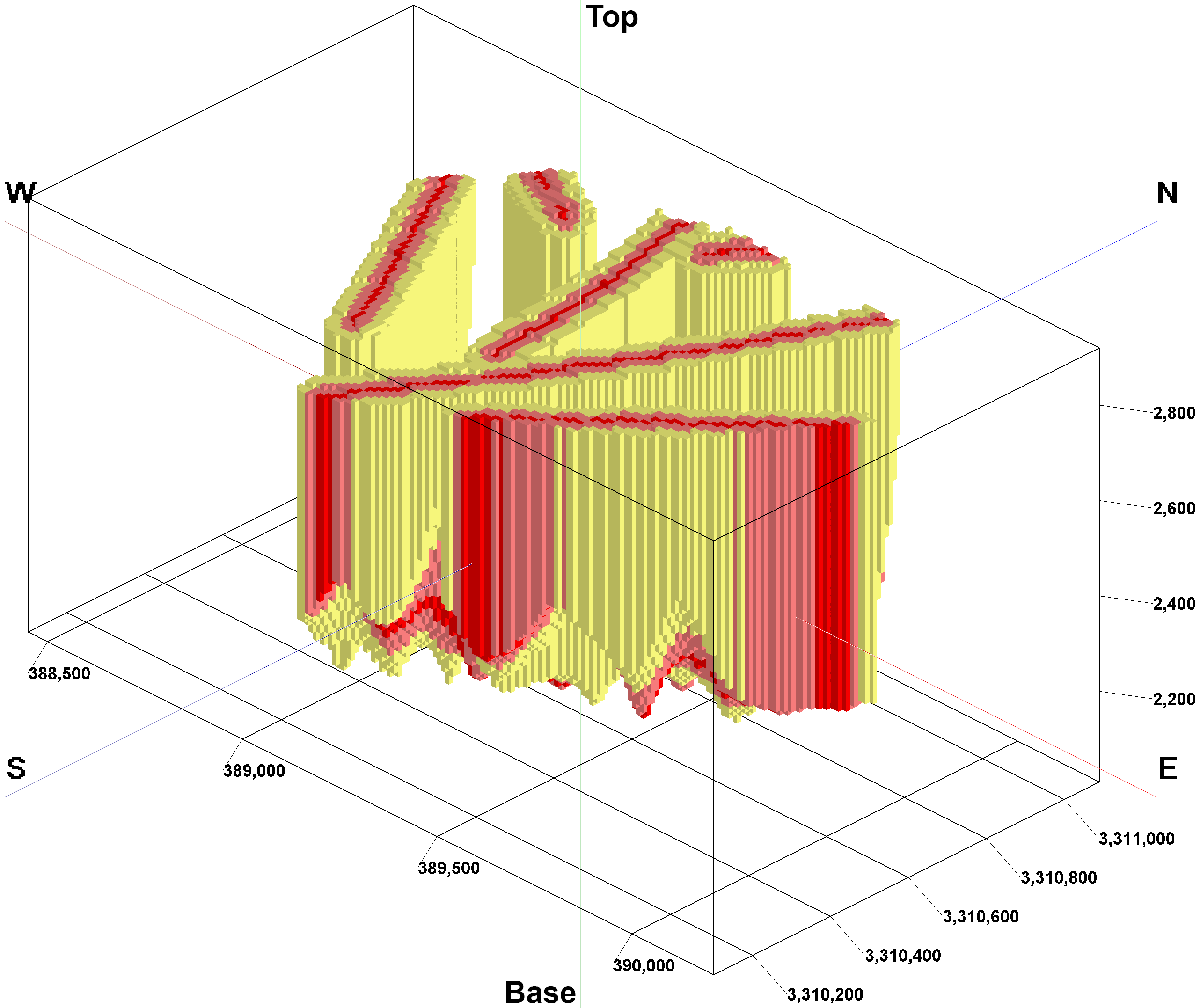}}
  \caption{Block model of the fault surfaces surrounded by two buffer zones in a radius of 25 and 50 m.}
  \label{fig13}
\end{figure}

We use the Cu concentration values in different intervals through the boreholes to create a primary 3D model of the target ore body concealed in depth using the inverse-distance anisotropic interpolation method which is later used as a training model. Moreover, we use Fe, Mo, and Zn concentration values for creating other geochemical evidential models. We present the descriptive statistics of the output models in Table 2 and the anomalous voxels in Fig. 14. The threshold used in each model for discriminating between the anomaly population and the background except for the Cu concentration, equals the 95th percentile. According to the cutoff grade of Cu in the ore deposits located in the neighborhood of our study area such as Sarcheshmeh \citep{Waterman1975}, the threshold for creating a binary model of the target ore body is considered 0.4\%. Based on this threshold, the target ore body occupies almost 3.5\% of the total modeling space, which can also be considered as the prior probability (Fig. 14a).

\begin{table}[]
  \caption{Descriptive statistics of the interpolated geochemical models.}
  \label{table2}
  \resizebox{\linewidth}{!}{%
  \begin{tabular}{llllllll}
    \hline\noalign{\smallskip}
    Element  & Minimum   & Maximum     & Mean      & Median     & Standard Deviation & Skewness & Kurtosis \\
    \noalign{\smallskip}\hline\noalign{\smallskip}
    Cu (\%)  & 0         & 1.9466      & 0.1758    & 0.1559     & 0.1095             & 2.8513   & 20.3918  \\
    Fe (ppm) & 1046.9552 & 227389.6406 & 50080.683 & 50797.1602 & 15825.3686         & 0.2302   & 2.9143   \\
    Mo (ppm) & 0.0155    & 2613.1301   & 187.9939  & 165.018    & 141.2469           & 2.023    & 10.4016  \\
    Zn (ppm) & 0.0327    & 27061.6719  & 339.8742  & 156.7078   & 726.8055           & 13.0335  & 273.5633 \\
    \noalign{\smallskip}\hline
  \end{tabular}}
\end{table}

\begin{figure}
  \centering
  \frame{\includegraphics[width=\linewidth]{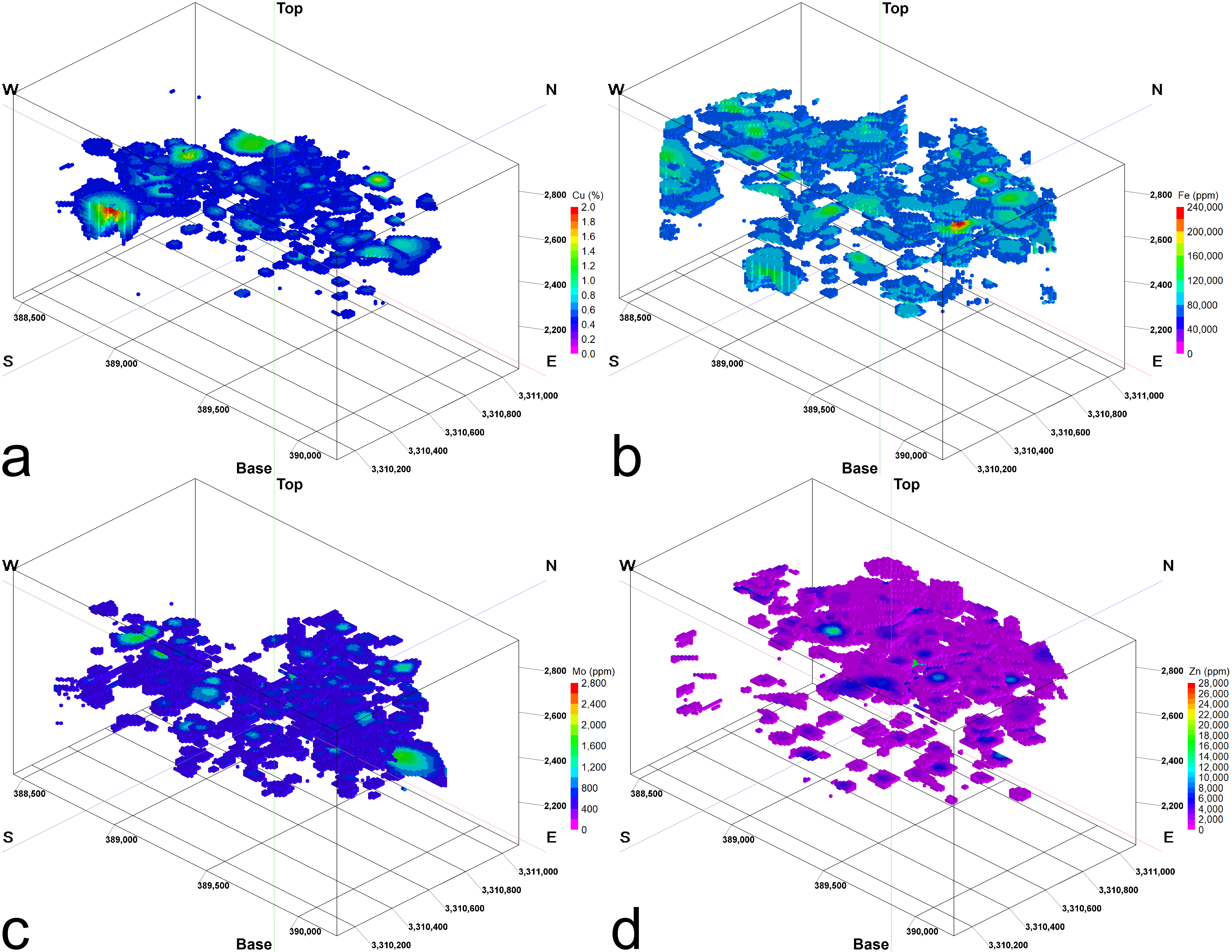}}
  \caption{Anomalous voxels of the 3D geochemical models created using the concentration values of a) Cu, b) Fe, c) Mo, and d) Zn.}
  \label{fig14}
\end{figure}

\subsection{Prospectivity modeling}
\label{sec5-2}
\subsubsection{Posterior probability model}
\label{sec5-2-1}
Based on the equations presented in section 4.4, the ordinary weights of evidence for geological evidential models are determined (Table 3). We consider every unit of lithology, alteration, and rock type data as a binary model in order to determine the weights of evidence, contrast, and studentized contrast. Those lithology and alteration units as well as different rock types and structural models which are not found in this table show negative contrast and are removed from the modeling process. One of the reasons can be the low number of occupied voxels, and hence low number of common voxels with the target ore body. More details can be found in the supplementary file. The fuzzy weight along with other important parameters are determined for each class of the geochemical evidential models. In Tables 4--6, we present the results for Fe, Mo, and Zn models, respectively. It is noteworthy that the classes with negative contrast are neglected during the modeling process. \\

\begin{table}[]
  \caption{Variation of the ordinary weights of evidence and other important parameters with different lithology and alteration units.}
  \label{table3}
  \begin{tabular}{lllll}
    \hline\noalign{\smallskip}
                 & W\textsuperscript{+}     & W\textsuperscript{-}      & Contrast & Studentized Contrast \\
    \noalign{\smallskip}\hline\noalign{\smallskip}
    Lithology    &        &         &          &                      \\
    Quartzolite  & 0.7046 & -0.0016 & 0.7062   & 5.0496               \\
    Alteration   &        &         &          &                      \\
    Calcitized   & 0.121  & -0.0026 & 0.1236   & 2.3717               \\
    Carbonatized & 0.6429 & -0.0033 & 0.6463   & 6.8551               \\
    Epidotized   & 2.2307 & -0.0001 & 2.2308   & 2.732                \\
    Potassic     & 0.1105 & -0.1611 & 0.2716   & 17.061               \\
    Silicified   & 0.9206 & -0.0118 & 0.9324   & 16.2444              \\
    \noalign{\smallskip}\hline
  \end{tabular}
\end{table}

\begin{table}[]
  \caption{Variation of the fuzzy weights of evidence and other important parameters with different classes of Fe concentration values.}
  \label{table4}
  \resizebox{\linewidth}{!}{%
  \begin{tabular}{llllllll}
    \hline\noalign{\smallskip}
    Lower Limit & Upper Limit & W\textsuperscript{+} & W\textsuperscript{-} & Contrast & Studentized Contrast & Fuzzy Contrast & Fuzzy Weight \\
    \noalign{\smallskip}\hline\noalign{\smallskip}
    Min        & 30086.0527 & -1.5855 & 0.0872  & -1.6728 & -31.5801 & 0.01   & 0.0863  \\
    30086.0527 & 38639.6172 & -0.1371 & 0.0142  & -0.1513 & -5.5568  & 0.6648 & -0.0116 \\
    38639.6172 & 43641.5    & -0.0218 & 0.0024  & -0.0242 & -0.9349  & 0.7551 & -0.0037 \\
    43641.5    & 47356.9648 & 0.0021  & -0.0002 & 0.0024  & 0.0923   & 0.7717 & 0.0004  \\
    47356.9648 & 50729.9414 & 0.0033  & -0.0004 & 0.0037  & 0.1437   & 0.7725 & 0.0006  \\
    50729.9414 & 54003.6758 & 0.0134  & -0.0015 & 0.0149  & 0.5815   & 0.7793 & 0.0027  \\
    54003.6758 & 57515.7734 & -0.028  & 0.0031  & -0.031  & -1.1927  & 0.7507 & -0.0046 \\
    57515.7734 & 61537.293  & -0.2199 & 0.0219  & -0.2418 & -8.5735  & 0.5916 & -0.0086 \\
    61537.293  & 67457.7891 & -0.0614 & 0.0066  & -0.068  & -2.5801  & 0.7259 & -0.0085 \\
    67457.7891 & Max        & 0.8286  & -0.1469 & 0.9755  & 51.4817  & 0.99   & 0.7728  \\
    \noalign{\smallskip}\hline
  \end{tabular}}
\end{table}

\begin{table}[]
  \caption{Variation of the fuzzy weights of evidence and other important parameters with different classes of Mo concentration values.}
  \label{table5}
  \resizebox{\linewidth}{!}{%
  \begin{tabular}{llllllll}
    \hline\noalign{\smallskip}
    Lower Limit & Upper Limit & W\textsuperscript{+} & W\textsuperscript{-} & Contrast & Studentized Contrast & Fuzzy Contrast & Fuzzy Weight \\
    \noalign{\smallskip}\hline\noalign{\smallskip}
    Min      & 37.036   & -1.3426 & 0.0811  & -1.4237 & -30.1982 & 0.1718 & 0.0634  \\
    37.036   & 74.6444  & -0.4013 & 0.0365  & -0.4378 & -14.3426 & 0.6106 & -0.0181 \\
    74.6444  & 108.0786 & -0.0999 & 0.0105  & -0.1104 & -4.1198  & 0.7543 & -0.0165 \\
    108.0786 & 137.7597 & -0.0856 & 0.0091  & -0.0947 & -3.556   & 0.7602 & -0.0148 \\
    137.7597 & 166.287  & -0.0341 & 0.0037  & -0.0378 & -1.4497  & 0.7808 & -0.0069 \\
    166.287  & 196.6305 & -0.0773 & 0.0082  & -0.0855 & -3.2213  & 0.7636 & -0.0137 \\
    196.6305 & 232.1037 & 0.1468  & -0.0176 & 0.1644  & 6.7928   & 0.8436 & 0.047   \\
    232.1037 & 280.1264 & 0.2393  & -0.0302 & 0.2695  & 11.548   & 0.87   & 0.093   \\
    280.1264 & 357.9801 & 0.244   & -0.0309 & 0.2749  & 11.801   & 0.8712 & 0.0958  \\
    357.9801 & Max      & 0.528   & -0.0783 & 0.6063  & 28.9463  & 0.9303 & 0.3223  \\
    \noalign{\smallskip}\hline
  \end{tabular}}
\end{table}

\begin{table}[]
  \caption{Variation of the fuzzy weights of evidence and other important parameters with different classes of Zn concentration values.}
  \label{table6}
  \resizebox{\linewidth}{!}{%
  \begin{tabular}{llllllll}
    \hline\noalign{\smallskip}
    Lower Limit & Upper Limit & W\textsuperscript{+} & W\textsuperscript{-} & Contrast & Studentized Contrast & Fuzzy Contrast & Fuzzy Weight \\
    \noalign{\smallskip}\hline\noalign{\smallskip}
    Min      & 67.9985  & -0.9715 & 0.0684  & -1.0399 & -26.2799 & 0.3132 & 0.0359  \\
    67.9985  & 84.6815  & -0.474  & 0.0417  & -0.5157 & -16.358  & 0.572  & -0.0125 \\
    84.6815  & 102.5373 & 0.1285  & -0.0153 & 0.1438  & 5.8967   & 0.8379 & 0.0394  \\
    102.5373 & 125.5783 & 0.274   & -0.0352 & 0.3093  & 13.4312  & 0.8789 & 0.1138  \\
    125.5783 & 155.639  & 0.2312  & -0.029  & 0.2602  & 11.1136  & 0.8678 & 0.0884  \\
    155.639  & 198.6275 & 0.1243  & -0.0147 & 0.139   & 5.6916   & 0.8366 & 0.0378  \\
    198.6275 & 267.4829 & 0.0086  & -0.001  & 0.0096  & 0.375    & 0.797  & 0.002   \\
    267.4829 & 394.8134 & 0.086   & -0.01   & 0.096   & 3.8698   & 0.8242 & 0.0239  \\
    394.8134 & 700.7863 & 0.0092  & -0.001  & 0.0103  & 0.4007   & 0.7972 & 0.0021  \\
    700.7863 & Max      & 0.07    & -0.0081 & 0.078   & 3.1242   & 0.8188 & 0.0186  \\
    \noalign{\smallskip}\hline
  \end{tabular}}
\end{table}

Based on the results presented in Tables 4--6, we plot the variations of contrast and fuzzy weight, and variance of contrast and studentized contrast with increasing the concentration values of the investigated elements shown in Fig. 15. \\

\begin{figure}
  \centering
  \includegraphics[width=\linewidth]{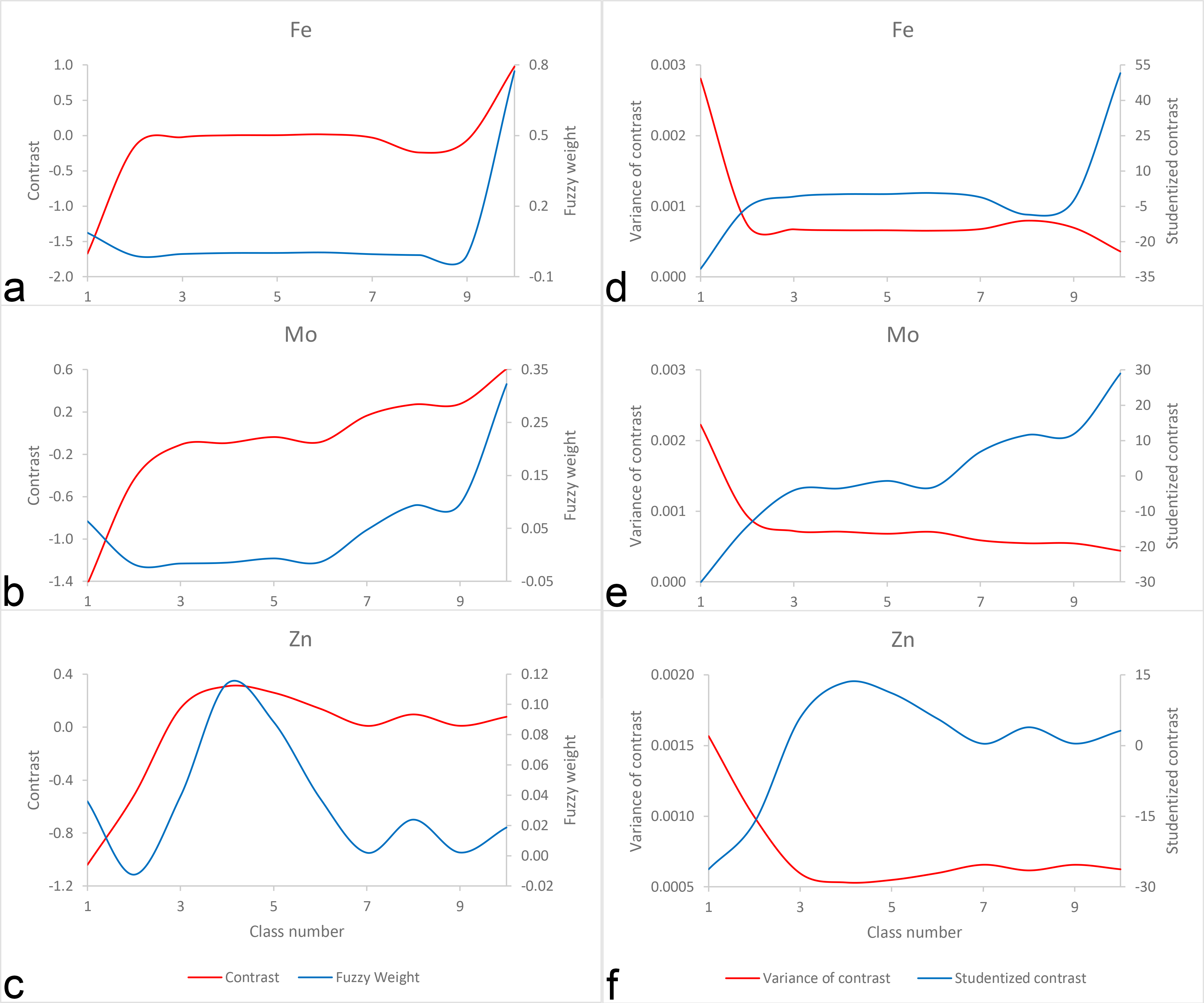}
  \caption{Variations of contrast and fuzzy weight with the concentration values of a) Fe, b) Mo, and c) Zn. Variations of variance of contrast and studentized contrast with the concentration values of d) Fe, e) Mo, and f) Zn.}
  \label{fig15}
\end{figure}

All the evidential models presented in Tables 3--6 are integrated as described in section 4.4, and a posterior probability model is the result. The C-V fractal model is used for classification and determining a proper threshold for separating anomalous voxels. According to the C-V chart presented in Fig. 16a, the voxels showing a posterior probability greater than 0.23 are considered as the certain anomaly. Also, the voxels showing a value greater than 0.15 and 0.09 can be considered as the probable and possible anomaly, respectively. In Fig. 16b, the classified posterior probability model and the voxels with values greater than 0.08 are shown which occupy nearly 3\% of the modeling space. \\

\begin{figure}
  \centering
  \frame{\includegraphics[width=\linewidth]{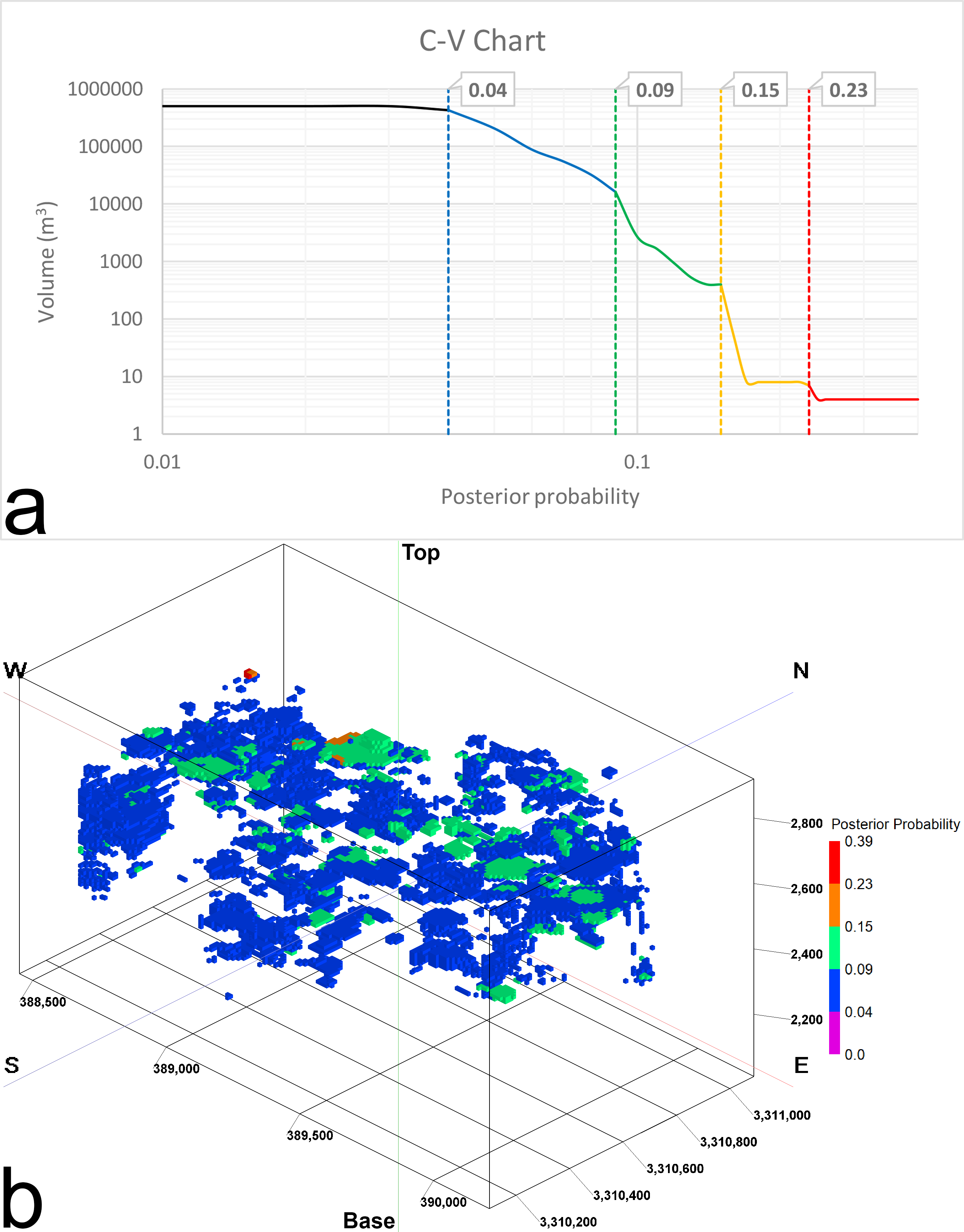}}
  \caption{a) C-V chart of the posterior probability of Cu mineralization in the modeling space. b) Classified anomalous voxels extracted based on the thresholds determined by the C-V chart.}
  \label{fig16}
\end{figure}

\subsubsection{Studentized posterior probability model}
\label{sec5-2-2}
The total variance of each voxel is given by summing the variances of the ordinary and fuzzy weights of evidence, which is considered as an estimate of the uncertainty associated with each voxel. Similar to the posterior probability model, we use the C-V chart for classification and separating the anomalous voxels. According to the relevant chart shown in Fig. 17a, the voxels with values greater than 4.9 are considered as a certain anomaly. Those voxels showing values greater than 3.8 and occupying nearly 3\% of the modeling space are shown in Fig. 17b.

\begin{figure}
  \centering
  \frame{\includegraphics[width=\linewidth]{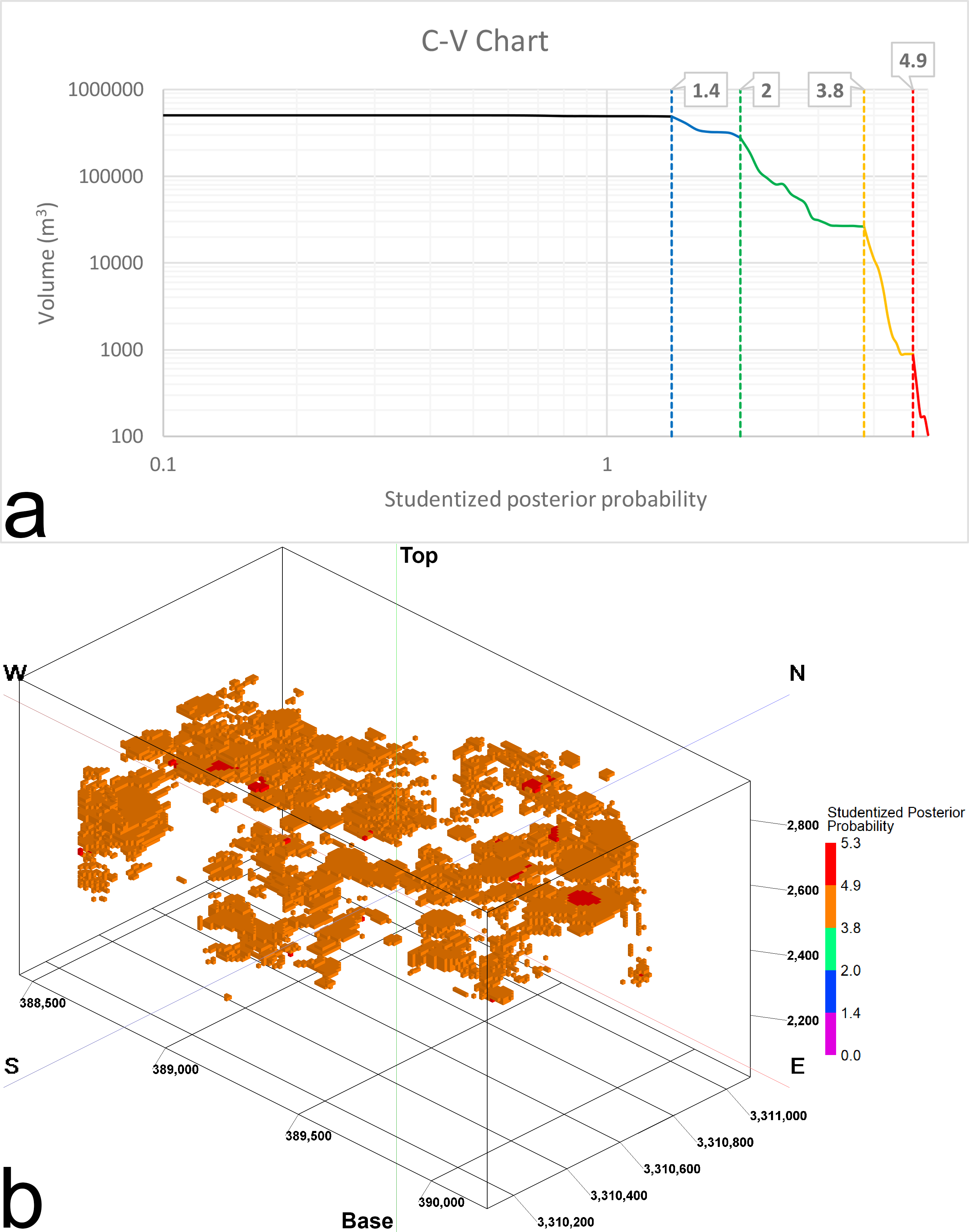}}
  \caption{a) C-V chart of the studentized posterior probability of Cu mineralization in the modeling space. b) Classified anomalous voxels extracted based on the thresholds determined by the C-V chart.}
  \label{fig17}
\end{figure}

\subsection{Validation}
\label{sec5-3}
In 2D mineral prospectivity mapping, a prediction-area plot is used in order to quantitatively validate the results obtained from prospectivity maps \citep{Yousefi2015}. We extend the application of this type of plots to three-dimensional space and call them prediction-volume (P-V) plots. In P-V plots, we show the cumulative percentage of predicted mineralization and the corresponding cumulative occupied volume, with respect to the total volume against the prospectivity values. Therefore, the prediction ability of a prospectivity model and its ability to delimit the modeling space for further exploration and drilling are evaluated in a scheme. The P-V plot shows a curve of the percentage (prediction rate) of known mineralization and a curve of the percentage of occupied volume corresponding to the classes of a prospectivity model. When an intersection point of the two curves is at a higher place, it portrays a small volume containing a large number of mineralization-bearing voxels. The comparison of prediction rates in the P-V plots shown in Fig. 18, indicates the importance of analyzing the predictability of the prospectivity models. We compared two prospectivity models including the posterior and studentized posterior probability models. It is noteworthy that for assigning probabilistic values to both models in terms of prospecting for Cu mineralization, and distribution of the voxel values between 0 and 1, they are transformed to a fuzzy space using a linear function.

\begin{figure}
  \centering
  \includegraphics[width=\linewidth]{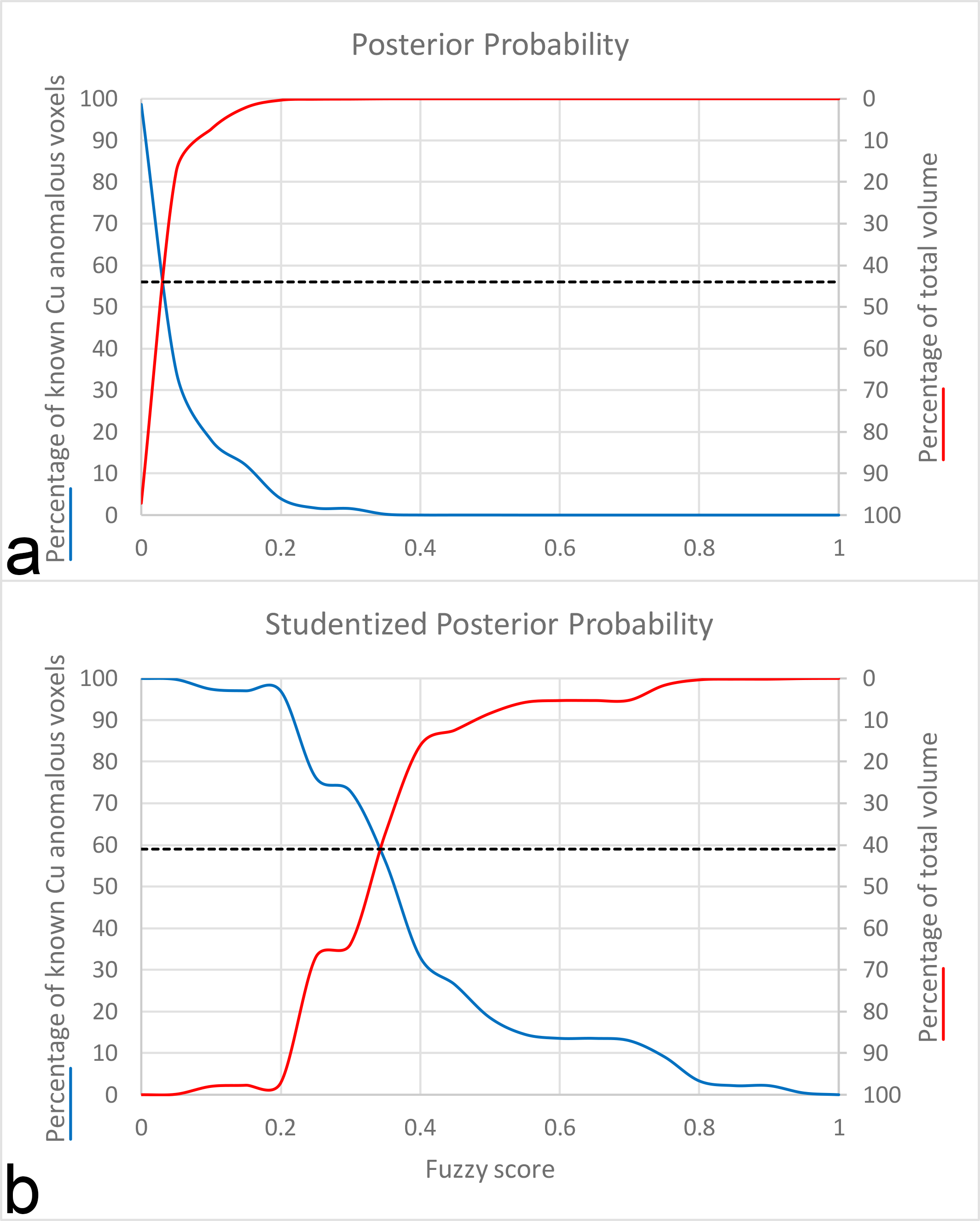}
  \caption{P-V plots for a) the posterior probability and b) studentized posterior probability prospectivity models.}
  \label{fig18}
\end{figure}

\section{Discussion}
\label{sec6}
Most of the study area is covered by alluvium, therefore the surface outcrop does not allow the lithology and alteration assemblages to be identified as detailed as the drill cores. In this study, we used two different types of drilling data including qualitative geological and quantitative geochemical data obtained from 113 boreholes through creating a 3D prospectivity model of Cu mineralization. The geological data comprise three types of data including lithology, alteration, and rock type. Among the lithology types, quartzolite units (i.e., intrusive rocks with a quartz content >90\%) show positive contrast and studentized contrast (Table 3) which were used as an input binary model to the modeling process. The quartzolite units which involve less than 1\% of the modeling space show a high studentized contrast indicating high correlation of this lithology type with Cu mineralization in the modeling space. Based on Figs. 11a and 12a, most of the modeling space is occupied by andesite which shows a contrast near zero and uncorrelated with Cu mineralization. The granodiorite units are in the second place in terms of volume percentage and show a negative contrast. According to Fig. 11a, high-grade Cu concentration is associated with stockwork veins mainly hosted by andesite which can be due to the high porosity and permeability of this lithology type caused by fractures. These probable fractures provide a path for hydrothermal fluids through ascending to the ground surface. In Fig. 11a, the cross-section $B-B'$ appears to show a vertical pipe hosted by andesite and characterized by elevated Cu concentration where the section $A-A'$ crosses section $B-B'$. On the other hand, granodiorite units can be considered barren or poorly mineralized. \\
Among the different types of alteration in the study area, calcitized, carbonatized, epidotized, potassic, and silicified units show positive contrast and studentized contrast (Table 3). These units were used as input binary models to the modeling process. According to the results, potassic and silicified units show the highest studentized contrast values, respectively. It is noteworthy that the contrast of silica alteration is much higher than the potassic alteration. This is compatible with the metallogenic model of the mineralization in the study area, because Cu mineralization is mostly observed along the silicified veins. The carbonatized units show a relatively high contrast, and according to the metallogenic model, malachite and azurite are considered as the main ore minerals. The epidotized and calcitized units are the last alteration units in terms of studentized contrast. The epidotized units occupy a small portion of the modeling space and the results are not reliable, although they show a high contrast value. The calcitized units show a low contrast value and then, a weak association with Cu mineralization in the modeling space. \\
According to Fig. 6c, the rocks located in the modeling space originate from five different sources, and none of them showed a positive contrast. Therefore, we did not use this type of geological data in the modeling process. We used surficial structural data through creating a 3D model of the fault surfaces due to lack of structural information in the drilling data. In addition to the fault surfaces which were converted into the fault blocks, two different buffer zones in a radius of 25 and 50 m were created surrounding the fault blocks. All the structural models show a negative contrast and were removed from the modeling process. This confirms the field observations implying that the Cu-bearing mineralization zones in the study area, are mainly associated with azurite and malachite stockwork veins and less associated with the faults. \\
Besides the qualitative geological data, we used the quantitative geochemical data consisting of the concentration values of three elements including Fe, Mo, and Zn. These elements are known as the key indicators of Cu mineralization in the study area. The interpolated model of the Cu concentration values used as the training model, ranges from 0 to nearly 2\% with an average of 0.17\%. The model is positively skewed and shows a high kurtosis value. Based on the descriptive statistics presented in Table 2, the evidential model of Fe concentration is the only model which shows a normal behavior with a skewness near zero and a kurtosis near 3 \citep{Montgomery2010}. The concentration values in this model range from nearly 0.1 to more than 22\% with an average of 5\%. The Mo model ranges from nearly 0.01 to more than 2600 ppm, showing an average of nearly 188 ppm. This model is positively skewed and shows a high kurtosis value. The Zn model which ranges from nearly 0.03 to more than 27000 ppm, is highly skewed and shows a very high kurtosis value.
As explained in section 4.4, to minimize the loss of information through converting continuous models into binary models, we determined the fuzzy weight of each class. Instead of using only a specific class of the continuous models, we used all those classes showing a positive contrast in the prospectivity modeling process. According to Tables 4--6, the highest studentized contrast values belong to the classes from Fe (51.4817), Mo (28.9463), and Zn (13.4312) models, respectively. Based on the graphs presented in Fig. 15, there is no linear relationship between contrast and fuzzy weight in continuous models, and a high contrast does not necessarily yield a high fuzzy weight. On the other hand, we see a clear inverse relationship between variance of contrast and studentized contrast. \\
We created two probability models including posterior and studentized posterior probability models for visualizing Cu mineralization potential in the target modeling space. It is clear that the studentized posterior probability model is more reliable due to the contribution of uncertainty through determining the anomalous voxels. According to Fig. 17b, where there are a low number of boreholes, the total variance or uncertainty shows a high value which gives a low studentized posterior probability. The intersection point in the P-V plot of the posterior probability model shows 56\% of the anomalous voxels in the interpolated model of Cu concentration have been predicted in 44\% of the total modeling space. The intersection point in the P-V plot corresponding to the studentized posterior model shows 59\% of the anomalous voxels in the interpolated model of Cu concentration have been predicted in 41\% of the total modeling space. The contribution of uncertainty in creating the studentized posterior probability model leads to a higher prediction rate compared to the posterior probability model. \\
According to Fig. 19, the ore body is divided into two parts. One is located in a shallow depth and southwest of the modeling space, while the other is located in a deeper level and northeast of the study area. The ore body located in the northeast is restricted to the fault surfaces, implying the younger age of the fault surfaces than the mineralization. The fault surfaces in the southwest are related to the mineralization and surrounded by anomalous voxels. These faults can be considered as the conduits through which the hydrothermal fluids have risen up near to the ground surface. The ore bodies and the difference in the depth of Cu mineralization are obvious and outlined in Fig. 20. \\

\begin{figure}
  \centering
  \frame{\includegraphics[width=\linewidth]{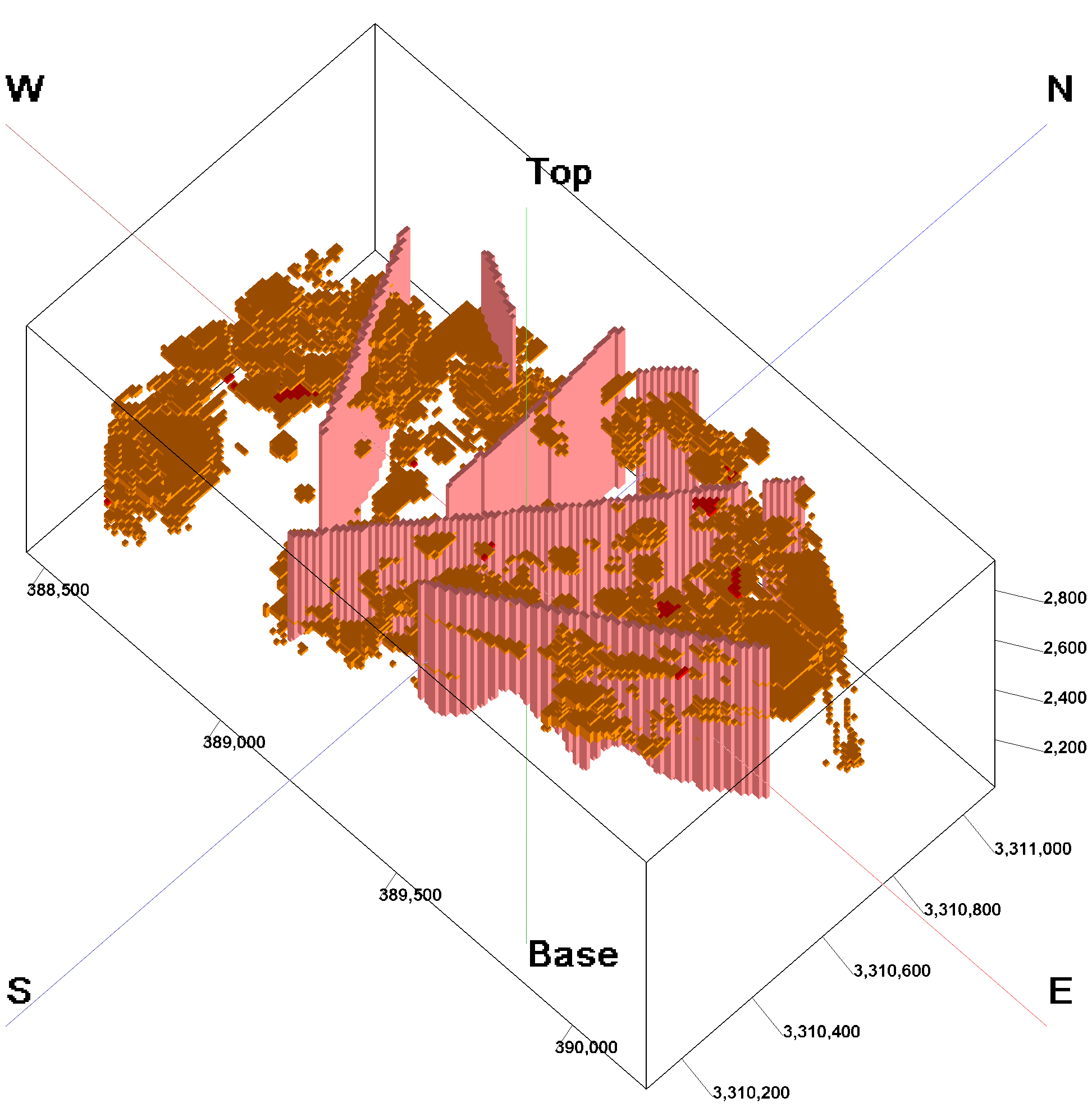}}
  \caption{Fault surfaces shown with the anomalous voxels of the studentized posterior probability model (Fig. 17b).}
  \label{fig19}
\end{figure}

\begin{figure}
  \centering
  \includegraphics[width=\linewidth]{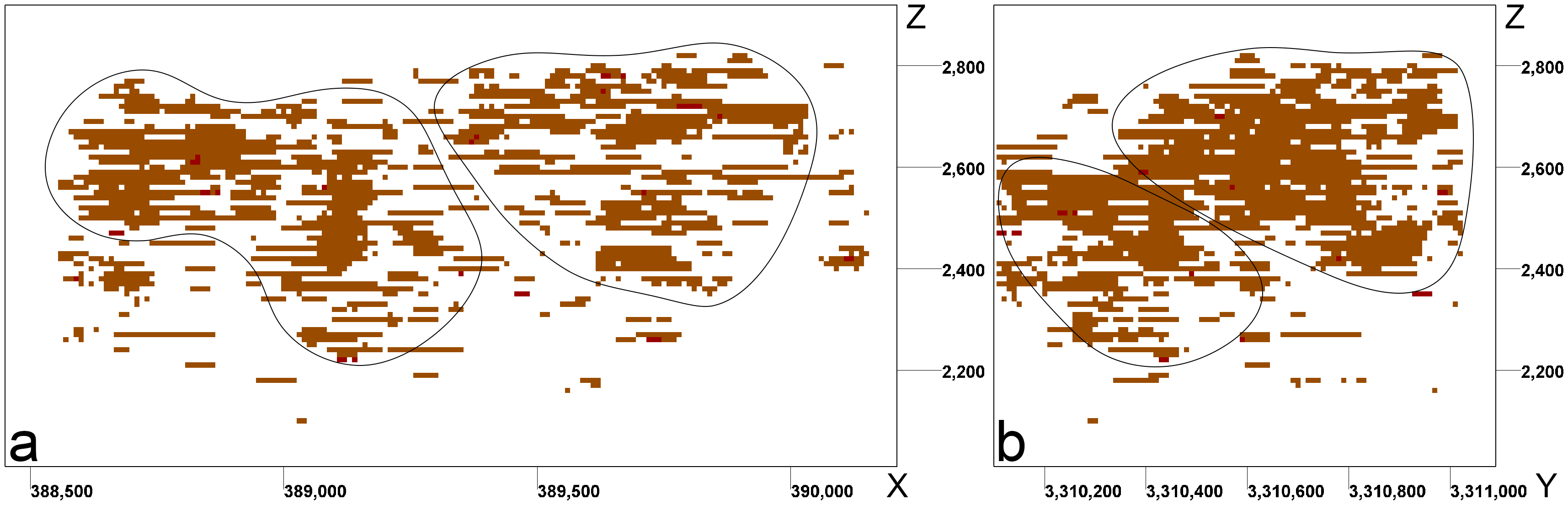}
  \caption{Side views of the studentized posterior probability model to the a) north, and b) west.}
  \label{fig20}
\end{figure}

We implemented the three-dimensional WofE modeling using Python scripts and release them as open-source software. The main advantage of these scripts is that they are simple and easy to understand for a user with basic knowledge of programming and can be extended for other types of problems.

\section{Conclusions}
In this study, we created a mineral prospectivity model based on 3D weights of evidence by integration of qualitative geological and quantitative geochemical borehole data located on a porphyry Cu deposit in southeast Iran. To minimize the loss of information, we determined the ordinary and fuzzy weights of evidence for discrete and continuous evidential models, respectively. The integration of various input evidential models provided two prospectivity models including posterior probability and studentized posterior probability models. The C-V fractal models were used through finding suitable thresholds for classifying and separating anomaly populations from the background. Although the studentized posterior probability model is more reliable due to the lower uncertainty, both the posterior and studentized posterior probability models show an acceptable prediction rate based on the P-V plots. The results show the efficiency of our framework in constructing different geometric models of a specific ore deposit concealed in depth. The proposed framework helps in determining key factors which control the mineralization in the modeling space, identifying potential mineralization, and improving the perception of ore genesis.
Among the lithological units, quartzolite shows a high correlation with Cu mineralization in the modeling space. Moreover, potassic, silica, carbonate, and propylitic alteration types are strongly associated with Cu mineralization. The fuzzy weights determined for continuous models show a high correlation between Cu and Fe concentration values, while Mo and Zn show an average correlation. According to the results of mineral prospectivity modeling, it is concluded that Cu mineralization is mainly associated with stockwork veins which confirms the field observations. The Cu mineralization is identified in two separate bodies located in different depths covered by a thin layer of regolith.
The hybrid application of ordinary and fuzzy weights of evidence yields a promising result for the future studies. The 3D mineral prospectivity modeling based on the weights of evidence is highly dependent on the primary model created using the concentration values of a specific target element. In future, we aim to use geostatistical methods for creating a more reliable model to consider as the primary target ore body, since we can provide an uncertainty model beside the model yielded by the interpolation process. In addition, we can decrease the uncertainty and increase the prediction rate in our posterior probability model by adding other exploration data such as geochemical and geophysical models and creating more input evidential models compatible to the metallogenic model of the study area.

\section*{Supplementary data}

The Python scripts for implementing the proposed framework and supplementary files including more details on the evidential models are available at \url{https://github.com/intelligent-exploration/3D_WofE}.

\begin{acknowledgements}
We acknowledge National Iranian Copper Industries Company for providing the data.
\end{acknowledgements}

\section*{Conflict of interest}

The authors declare that they have no conflict of interest.

\bibliographystyle{spbasic}      
\bibliography{References}   

\end{document}